\documentclass[journal,article,accept,moreauthors,pdftex,12pt,a4paper]{mdpi} 
\usepackage{pdflscape}
\usepackage{rotating}
\usepackage{amsmath}
\usepackage{amssymb}
\usepackage{graphicx}
\usepackage{caption}
\usepackage{subcaption}
\usepackage{ifoddpage}
\usepackage[T1]{fontenc}
\usepackage[utf8]{inputenc}
\usepackage{tabularx,ragged2e,booktabs,caption}
\newcolumntype{C}[1]{>{\Centering}m{#1}}

\lastpage{x}
\doinum{10.3390/------}
\pubvolume{xx}
\pubyear{2013}
\history{Received: xx / Accepted: xx / Published: xx}
\newcommand{\Process}{\mathcal{P}}

\newcommand{\MeasAlphabet}  {\mathcal{A}}
\newcommand{\MeasSymbol}   { {X} }
\newcommand{\meassymbol}   { {x} }
\newcommand{\MeasSymbols}[2]{ \MeasSymbol_{#1:#2} }
\newcommand{\meassymbols}[2] { \meassymbol_{#1:#2} }

\newcommand{\Past} { \MeasSymbols{}{0} }
\newcommand{\past} { \meassymbols{}{0} }

\newcommand{\Future} { \MeasSymbols{0}{} }
\newcommand{\CausalState}   { \mathcal{S} }

\newcommand{\Prob}      {\Pr} % use standard command

\newcommand{\Cmu}       {C_\mu}
\newcommand{\hmu}       {h_\mu}
\newcommand{\EE}        {{\bf E}}

\newcommand{\PC}        {\chi}

\newcommand{\ProcessAlphabet}   {\MeasAlphabet}

\newcommand{\forward}{+}
\newcommand{\reverse}{-}
\newcommand{\forwardreverse}{\pm} % \pm

\newcommand{\FutureCausalState} { {\CausalState}^{\forward} }

\newcommand{\PastCausalState}   { {\CausalState}^{\reverse} }

\newcommand{\lastindex}[2]{
  \edef\tempa{0}
  \edef\tempb{#2}
  \ifx\tempa\tempb
    \edef\tempc{#1}
  \else
    \edef\tempa{0}
    \edef\tempb{#1}
    \ifx\tempa\tempb
      \edef\tempc{#2}
    \else
      \edef\tempc{#1+#2}
    \fi
  \fi
  \tempc
}

\newcommand{\rmu}{r_\mu}
\newcommand{\bmu}{b_\mu}
\newcommand{\qmu}{q_\mu}

\newcommand{\sigmamu}{\sigma_\mu}

\newcommand{\CSjoint}[1][,]{
   \edef\tempa{:}
   \edef\tempb{#1}
   \ifx\tempa\tempb
      \ensuremath{\FutureCausalState\!#1\PastCausalState}
   \else
      \ensuremath{\FutureCausalState#1\PastCausalState}
   \fi
}

\newif\ifpm
\edef\tempa{\forwardreverse}
\edef\tempb{\pm}
\ifx\tempa\tempb
   \pmfalse
\else
   \pmtrue
\fi

\newcommand{\Abet}{\ProcessAlphabet}
\newcommand{\MS}{\MeasSymbol}
\newcommand{\ms}{\meassymbol}

\Title{Information Anatomy of Stochastic Equilibria}
\Author{Sarah Marzen $^{1*}$ and James P. Crutchfield $^{2*}$}
\address{%
$^{1}$ Department of Physics and Redwood Center for Theoretical Neuroscience,
  University of California at Berkeley, Berkeley, California 94720 USA\\
$^{2}$ Complexity Sciences Center, Department of Physics,
  University of California at Davis, One Shields Avenue, Davis,
  California 95616 USA}

\corres[2]{smarzen@berkeley.edu or chaos@ucdavis.edu}

\abstract{A stochastic nonlinear dynamical system generates information, as
measured by its entropy rate. Some---the ephemeral information---is
dissipated and some---the bound information---is actively stored and so affects
future behavior. We derive analytic expressions for the ephemeral and bound informations in the limit of small-time discretization for two classical systems that exhibit dynamical equilibria: first-order Langevin equations (i)
where the drift is the gradient of a potential function and the diffusion
matrix is invertible and (ii) with a linear drift term (Ornstein-Uhlenbeck) but
a noninvertible diffusion matrix. In both cases, the bound information is
sensitive only to the drift, while the ephemeral information is sensitive only
to the diffusion matrix and not to the drift. Notably, this information anatomy
changes discontinuously as any of the diffusion coefficients vanishes,
indicating that it is very sensitive to the noise structure. We then calculate
the information anatomy of the stochastic cusp catastrophe and of particles
diffusing in a heat bath in the overdamped limit, both examples of stochastic
gradient descent on a potential landscape. Finally, we use our methods to
calculate and compare approximations for the so-called time-local predictive
information for adaptive agents.
  }


\begin{document}

%%%%%%%%%%%%%%%%%%%%%%%%%%%%%%%%%%%%%%%%%%

\vspace{0.2in}
\section{Introduction}

If we track the position of a particle diffusing on an unchanging potential
long enough, we can estimate the probability of observing a sequence of
positions \cite{Walt82a}. From that, we can quantitatively answer questions
about the process's behavior using a range of information statistics:
\begin{itemize}
\setlength{\topsep}{-5pt}
\setlength{\itemsep}{-2pt}
\setlength{\parsep}{0pt}
\setlength{\labelwidth}{5pt}
\setlength{\itemindent}{0pt}
\item How random is it? The \textit{entropy rate $\hmu$}, which is the
	uncertainty in the present observation conditioned on all past
	observations \cite{Cove06a}.
\item What must be remembered about the past in order to optimally predict the
	future? The \textit{causal states}, which are groupings of pasts that
	lead to the same probability distribution over future trajectories
	\cite{Crut88a,Shal98a}.
\item How much memory is required to store these causal states? The
	\textit{statistical complexity $\Cmu$}, or the entropy of the
	causal states \cite{Crut88a}. % it sounded weird to say "information the causal states store", plus that's not precise enough for people to be like, oh okay, now i know what you're calculating.  you mentioned store in the question so i just changed that description back to the technical definition.
\item How much of the future is predictable from the past? The
	\textit{excess entropy $\EE$}, which is the mutual information between the
	past and the future \cite{Crut01a}.
\item How much of the generated information $(h_{\mu})$ is relevant to
	predicting the future? The \textit{bound information $b_{\mu}$}, which is
	the mutual information between the present and future observations
	conditioned on all past observations \cite{Jame11a}.
\item How much of the generated information is useless---neither affects future
	behavior nor contains information about the past? The
	\textit{ephemeral information} $r_{\mu}$, which is the uncertainty in the
	present observation conditioned on all past and future observations
	\cite{Jame11a}.
\end{itemize}
These informational quantities cannot be derived from a dynamical phase diagram
in general, so we see them as providing a complementary view of a
process's structure and behavior.

In applications, such informational characterizations of a time series are
useful for monitoring good sensory coding \cite{Palm13a}, cognitive modalities
\cite{Beer14a}, and brain coherence \cite{Tono98a}, hidden Markov model
structural inference \cite{Stre13a}, action policies of autonomous agents
\cite{Sato04a,Mart13a}, structure in disordered materials
\cite{Varn02a,Varn12a}, dynamical phase transitions \cite{Crut89e,Tche13a}, and
intrinsic information processing in deterministic chaos \cite{Atma91a,Jame13a}
and cellular automata \cite{Lizi10a,Flec11a}.

Here, we focus on continuous stochastic nonlinear dynamical systems, the theory
for which has a long and venerable history, has met with a number of successful
predictions, and has identified a number of principles describing how noise
interacts with nonlinearity \cite{Moss89a}. For nonlinear systems transitioning
to chaos, to take just one example, noise plays the role of a ``disordering''
field, just as the magnetic field is an ordering field for spin systems at
critical transitions \cite{Shra81,Crut81}. Though their history substantially
predates that of the wide range of complex systems just cited, relatively fewer
analyses of their information processing components---their
\emph{information anatomy}---have been carried out.  As a start, we
demonstrate how to calculate the quantities above for continuous-time,
continuous-state stochastic nonlinear systems exhibiting dynamical equilibria,
yielding intuition for the properties these measures capture in simpler, and
perhaps more familiar, physical models.

Throughout, we focus on a ubiquitous and simple nonlinear generative model:
stochastic gradient descent or, in other words, diffusion on a potential
surface. We assume infinite precision in our observation of the state space.
The first calculation assumes that the diffusion matrix is invertible; the
second assumes that the drift term is linear but allows for a noninvertible
diffusion matrix. All calculations assume that the time between measurements is
nonzero, but arbitrarily small.

To get started, background is given in Section \ref{sec:Background}. Results
are presented in Section \ref{sec:first-order} and stated more succinctly in
Table \ref{tab:title}. To illustrate how to apply those formulae, we calculate
the information anatomy of the stochastic cusp catastrophe in Section \ref{sec:cusp_catastrophe} and coupled particles diffusing in a heat bath in Section \ref{sec:particles_heatbath}.

We provide a suite of appendices that are home to technical details necessary
for completeness, but that would otherwise distract. Several appendices also
draw out implications of information anatomy analysis. In particular, Appendix
\ref{sec:Markov} shows that the information anatomy of a Markovian system
requires looking only one time step into the future and past, as expected from
a similar calculation in \cite{Jame11a}. Appendix \ref{sec:causal_states}
establishes that the causal states of a first-order Langevin equation are
isomorphic to the present position.  Appendix \ref{sec:Greens} justifies why,
given an infinitesimal time resolution $\tau$, the conditional entropy of the
measurement at a future time step given the present measurement can be
approximated arbitrarily well by using a linearized drift term when the
diffusion matrix is invertible.  Appendix \ref{sec:linearLangevin} then
demonstrates that the entropy of the Green's function of a linear Langevin
equation with a noninvertible diffusion matrix differs from that when the
diffusion matrix is invertible.  Finally, Appendix \ref{sec:TiPi} applies the
formulae in Appendices \ref{sec:Markov}-\ref{sec:Greens} to explore estimates
of the time-local predictive information and related alternatives, used as
optimization principles to choose action policies for adaptive autonomous
agents \cite{Mart13a}.

%%%%%%%%%%%%%%%%%%%%%%%%%%%%%%%%%%%%%%%%%%

\section{Background}
\label{sec:Background}

Let's first recall the information anatomy analysis of discrete-time,
discrete-state processes introduced in \cite{Jame11a}. The main object of study
is a process $\Process$: the list of all of a system's behaviors or realizations
$\{ \ldots \ms_{-2}, \ms_{-1}, \ms_{0}, \ms_{1}, \ldots \}$ and their
probabilities $\Prob(\ldots \MS_{-2}, \MS_{-1}, \MS_{0}, \MS_{1}, \ldots)$.
We denote a contiguous chain of random variables as
$\MS_{0:L} = \MS_0 \MS_1 \cdots \MS_{L-1}$. We assume the process is
ergodic and stationary---$\Prob(\MS_{0:L}) = \Prob(\MS_{t:L+t})$ for all
$t \in \mathbb{Z}$---and the measurement symbols range over a
finite alphabet: $\ms \in \Abet$. In this setting, the \emph{present} $\MS_0$
is the random variable measured at $t = 0$, the \emph{past} is the
chain $\MS_{:0} = \ldots \MS_{-2} \MS_{-1}$ leading up the present, and the
\emph{future} is the chain following the present
$\MS_{1:} = \MS_1 \MS_2 \cdots$.

Shannon's various information quantities---entropy, conditional entropy, mutual
information, and the like---when applied to time series are functions of the joint distributions $\Prob(\MS_{0:L})$. Importantly, they define an algebra of information measures for a given set of random variables \cite{Yeun08a}. Ref. \cite{Jame11a} used this to show that the past and future partition the single-measurement entropy
$H(\MS_0)$ into several measure-theoretic atoms. These include the
\emph{ephemeral information}:
\begin{align*}
\rmu = H[X_0|X_{:0},X_{1:}] ~,
\end{align*}
which measures the uncertainty of the present knowing the past and future;
the \emph{bound information}:
\begin{align*}
\bmu = I[X_0;X_{1:}|X_{:0}] ~,
\end{align*}
which is the information shared between present and future conditioned on past;
and the \emph{enigmatic information}:
\begin{align*}
\qmu = I[X_0;X_{:0};X_{1:}] ~,
\end{align*}
which is the co-information between past, present, and future.

For a stationary time series, the bound information is also the
shared information between present and past conditioned on the future:
\begin{align*}
\bmu = I[X_0;X_{:0}|X_{1:}].
\end{align*}
One can also consider the amount of predictable information not captured by the present:
\begin{align*}
\sigmamu = I[X_{:0};X_{1:}|X_0].
\end{align*}
which is called the \emph{elusive} information. It measures the amount of
past-future correlation not contained in the present. It is nonzero if the
process has ``hidden states'' and is therefore quite sensitive to how the state
space is ``observed'' or coarse-grained.

The total information in the future predictable from the past (or vice versa)
is the \emph{excess entropy}:
\begin{align*}
\EE = I[X_{:1};X_{1:}] = \bmu + \sigmamu + \qmu ~.
\end{align*}
The process's Shannon entropy rate $\hmu$ can also be written as a sum of
atoms:
\begin{align*}
\hmu = H[X_0|X_{:0}] = \rmu + \bmu ~.
\end{align*}
Thus, a portion of the information ($\hmu$) a process spontaneously generates is thrown away ($\rmu$) and a portion is actively stored ($\bmu$). Putting
these observations together gives the information anatomy of a single measurement:
\begin{align*}
H[X_0] = q_{\mu} + 2b_{\mu} + r_{\mu} ~.
\end{align*}
These quantities were originally defined for stationary processes, but
easily carry over to a nonstationary process of finite Markov order. (See
Appendix \ref{sec:Markov}.)

The burden of the following is to analyze the limit from the discrete-time,
discrete-value processes just discussed to continuous-time,
continuous-value processes. Suppose that
observations are made at very small intervals of duration $\tau$. Then the
observation at time $t_n = n \tau$ is now labeled $X_{n\tau}$. Rather than
entropy or mutual information per observed symbol, we define an entropy or
mutual information per elapsed time; that is, informational rates. A step in
that direction is to normalize the information measures defined above by the
observation interval:
\begin{align*}
\rmu(\tau)     &= H[X_0|X_{:0},X_{\tau:}]  / \tau ~,\\
\bmu(\tau)     &= I[X_{\tau:};X_0|X_{:0}]  / \tau ~,\\
\qmu(\tau)     &= I[X_{:0};X_0;X_{\tau:}]  / \tau ~,\\
\sigmamu(\tau) &= I[X_{:0};X_{1:}|X_0] / \tau ~,~\text{and}\\
H_0(\tau)      &= H[X_0] / \tau ~.
\end{align*}
In doing this, terms of order $\tau$ or higher are ignored. These definitions
then lead to a familiar $\tau$-entropy rate using a discrete-time,
continuous-value treatment \cite{Gasp93a,Gasp05a,Cove06a}:
\begin{align*}
\hmu (\tau) = H[X_0|X_{:0}] / \tau ~.
\end{align*}
More natural definitions of these quantities might involve a fully
continuous-time development that avoids the $\log\tau$ divergences of the $\tau$ entropy
rate \cite{Leco07a}, but we leave this for future research.

\begin{figure}
\checkoddpage
  \edef\side{\ifoddpage l\else r\fi}%
  \makebox[\textwidth][\side]{
\begin{subfigure}[h]{.5\linewidth}
  \includegraphics[width=\textwidth]{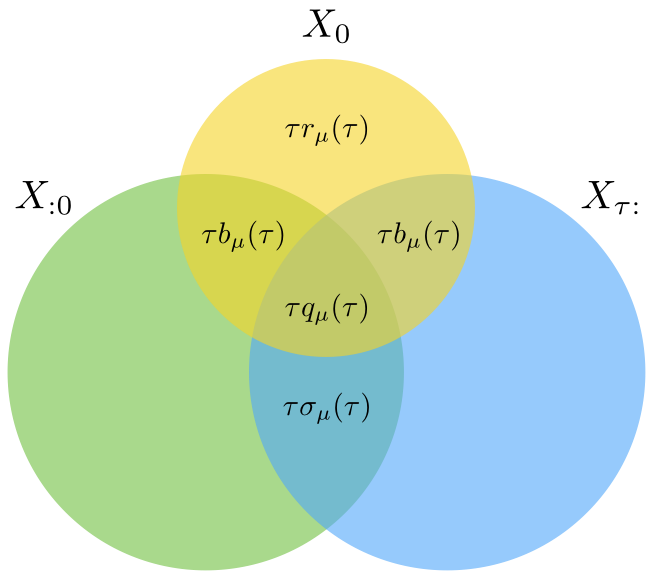}
  \caption{General information diagram for the anatomy of a single
	observation $\MS_0$.
  }
  \label{fig1A}
\end{subfigure}

\begin{subfigure}[h]{.5\linewidth}
  \includegraphics[width=\textwidth]{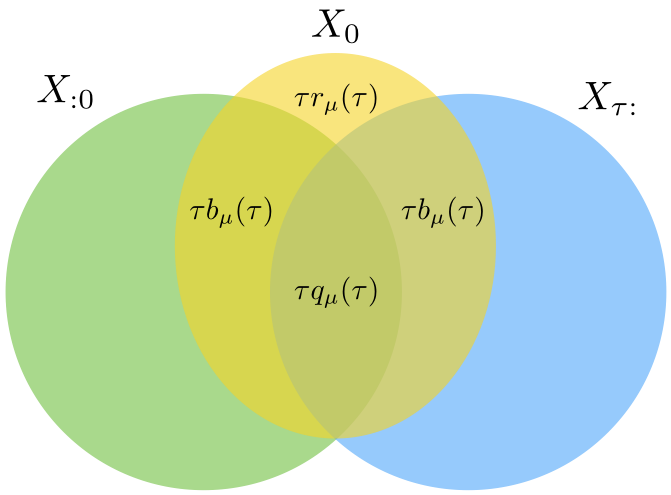}
  \caption{Information diagram for the anatomy of a Markovian process,
	in which the present $\MS_0$ causally shields the past from future.
	The elusive information $\sigmamu(\tau)$ vanishes.
  }
  \label{fig1B}
\end{subfigure}
}
\caption{Stationary continuous-time series information diagrams.}
\label{fig1}
\end{figure}

Figures \ref{fig1}\subref{fig1A} and \ref{fig1}\subref{fig1B} give \emph{information diagrams} that illustrate the algebra of the information measure atoms just defined. There, the entropy of a set is the sum of the entropy of its atoms. This reveals several useful linear dependencies that were
originally noted in \cite{Jame11a}:
\begin{align*}
H_0(\tau)  &= \rmu(\tau) + 2 \bmu (\tau) + \qmu (\tau) ~,\\
\hmu(\tau) &= \rmu(\tau) + \bmu (\tau), ~\text{and}\\
\EE / \tau  &= \qmu(\tau) + \sigmamu (\tau) + \bmu (\tau) ~.
\end{align*}
For a Markovian process, as illustrated in Figure \ref{fig1}\subref{fig1B},
the elusive information vanishes:
\begin{align*}
\sigmamu (\tau) = 0 ~.
\end{align*}
Therefore, in this case, if we find expressions for $H_0(\tau)$,
$\hmu (\tau)$, and $\bmu (\tau)$, then we can find $\rmu (\tau)$,
$\qmu (\tau)$, and $\EE$ via:
\begin{align}
\qmu (\tau) &= H_0(\tau)- \hmu (\tau) - \bmu (\tau) ~, \label{eq:qmu} \\
\rmu (\tau) &= \hmu (\tau) - \bmu (\tau) ~, \text{and} \label{eq:rmu} \\
\EE / \tau  &= H_0(\tau) - \hmu (\tau) ~. \label{eq:EMarkov}
\end{align}

%%%%%%%%%%%%%%%%%%%%%%%%%%%%%%%%%%%%%%%%%%

\section{Information Anatomy of Stochastic Dynamical Systems}
\label{sec:first-order}

To determine a process's information anatomy one must calculate entropies and
conditional entropies of the joint probability distribution of the entire
past, the present, and the entire future. In the general case, this is
challenging. However, since first-order Langevin equations we consider are 
Markovian, we have:
\begin{align}
\tau \hmu (\tau) &= H[X_{\tau}|X_{0}] ~\text{and} \label{eq:1main} \\
\tau \bmu (\tau) &= I[X_{\tau};X_0]-I[X_{\tau};X_{-\tau}] ~.
	\label{eq:2main}
\end{align}
(Appendix \ref{sec:Markov} provides the derivation.) Therefore, to calculate a
Markovian process's information anatomy, we only need the joint probability
distribution of three successive measurements instead of the joint
probability distribution of the present and semi-infinite past and future.
To further
simplify the calculation of conditional entropies, we assume that $\tau$ is
small enough that the entropy of the Green's function---i.e., the transition
probabilities $P(x',t+\tau|x,t)$---is well approximated by the entropy of a
corresponding Gaussian. This is exactly true for a linear Langevin equation.
For a nonlinear Langevin equation, the Gaussian approximation is valid in the limit of
infinitesimal $\tau$ for a set whose measure can be made arbitrarily close to $1$. (Appendix \ref{sec:Greens}
calculates small-$\tau$ approximations for the variance of this Gaussian.)
We do \textit{not} approximate the stationary distribution of a nonlinear Langevin equation by a Gaussian, however, and that means that the joint probability distribution over successive measurements is in general highly
non-Gaussian.

Appendix \ref{sec:causal_states} shows that, for first-order Langevin
dynamics, the single-measurement entropy $H[X_0]$ is the process's statistical
complexity $\Cmu$ \cite{Crut88a,Shal98a}. The result is that the information
anatomy analysis decomposes this causal-state information into:
\begin{itemize}
\setlength{\topsep}{0pt}
\setlength{\itemsep}{0pt}
\setlength{\parsep}{0pt}
\setlength{\labelwidth}{5pt}
\setlength{\itemindent}{0pt}
\item that useful for prediction or retrodiction beyond the information
	provided by the causal states at the previous time step---the
	bound information $\bmu$;
\item that useful for both prediction and retrodiction---the co-information
	$\qmu$; and
\item that useless for both prediction and retrodiction---the ephemeral
	information rate $\rmu$.
\end{itemize}
This is a similar but finer $\Cmu$ decomposition than considered in
\cite{Crut08b}. There, and more generally, $\Cmu = \EE + \PC$. That is, the
state information consists of that shared with the future ($\EE$) and
information not shared with the future but that must be stored to implement
optimal prediction---the \emph{crypticity} $\PC$ \cite{Crut08a}. Together with
these observations, Eqn. \ref{eq:EMarkov} reminds us that $\PC = \hmu$ for
Markov processes, as originally noted for finite-range one-dimensional spin
systems \cite{Crut97a}.

\subsection{Nonlinear Langevin Dynamics}
\label{sec:first-ordernonlinearLangevin}

\begin{figure}
\checkoddpage
  \edef\side{\ifoddpage l\else r\fi}%
  \makebox[\textwidth][\side]{
\begin{subfigure}[h]{.5\linewidth}
\includegraphics[width=\textwidth]{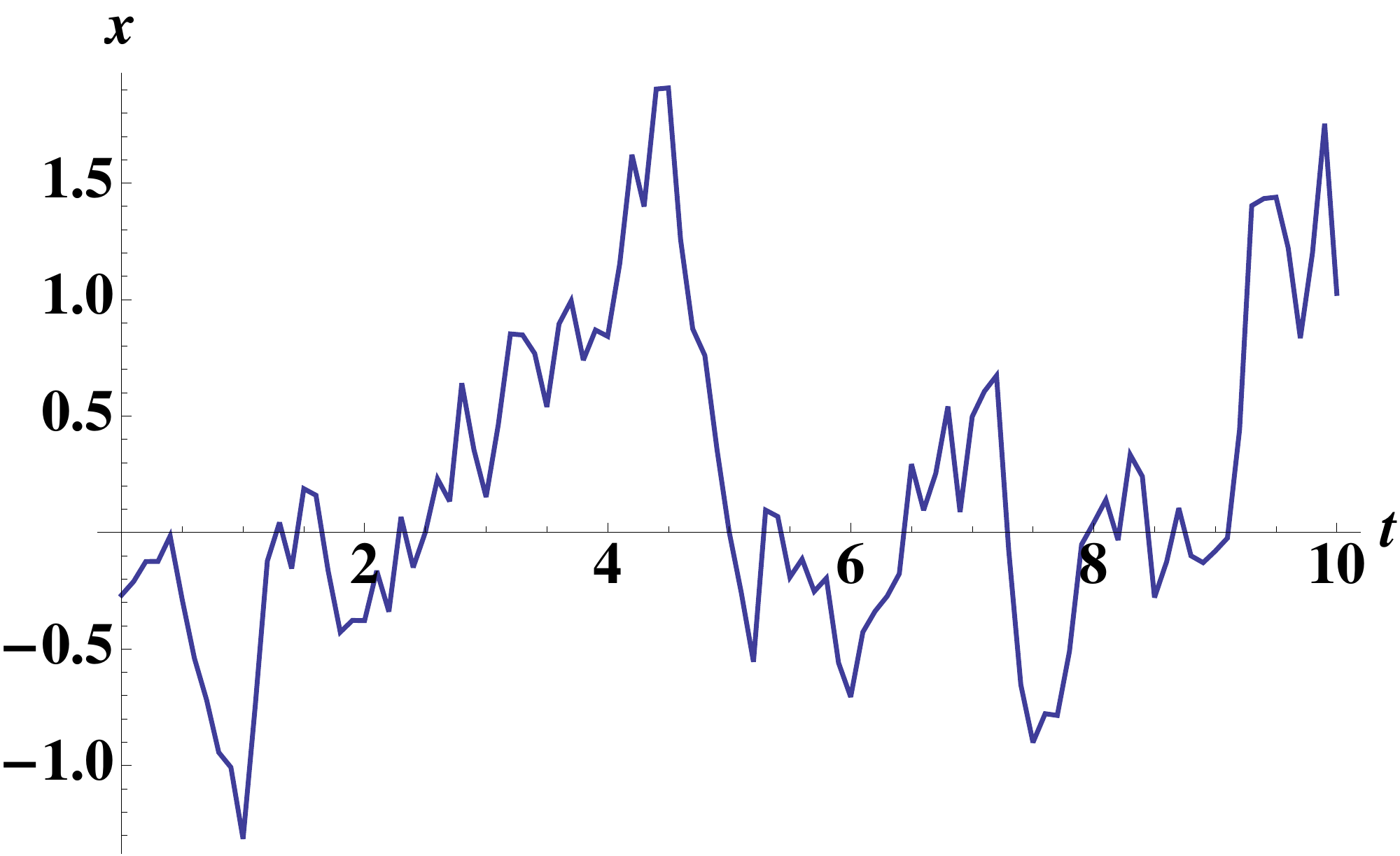}
  \caption{A finite-time trajectory $x(t)$ followed by the diffusing particle.
  }
\label{fig1A2}
\end{subfigure}

\begin{subfigure}[h]{.5\linewidth}
\includegraphics[width=\textwidth]{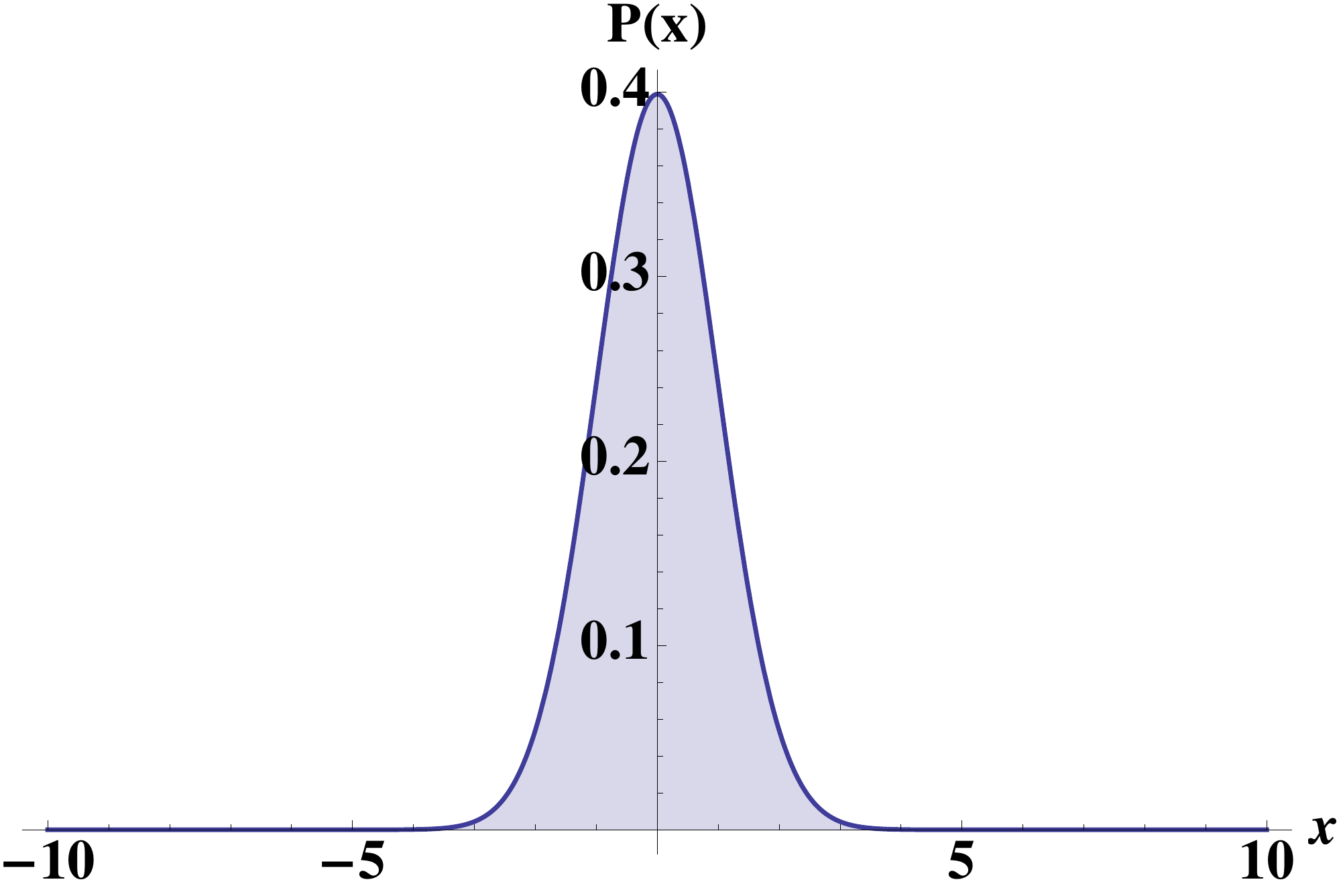}
  \caption{Equilibrium probability distribution $\rho_{eq}(x)$,
  calculated as a normalized histogram of particle positions.
  }
\label{fig1B2}
\end{subfigure}
}
\centering
\begin{subfigure}[h]{.8\linewidth}
\includegraphics[width=\textwidth]{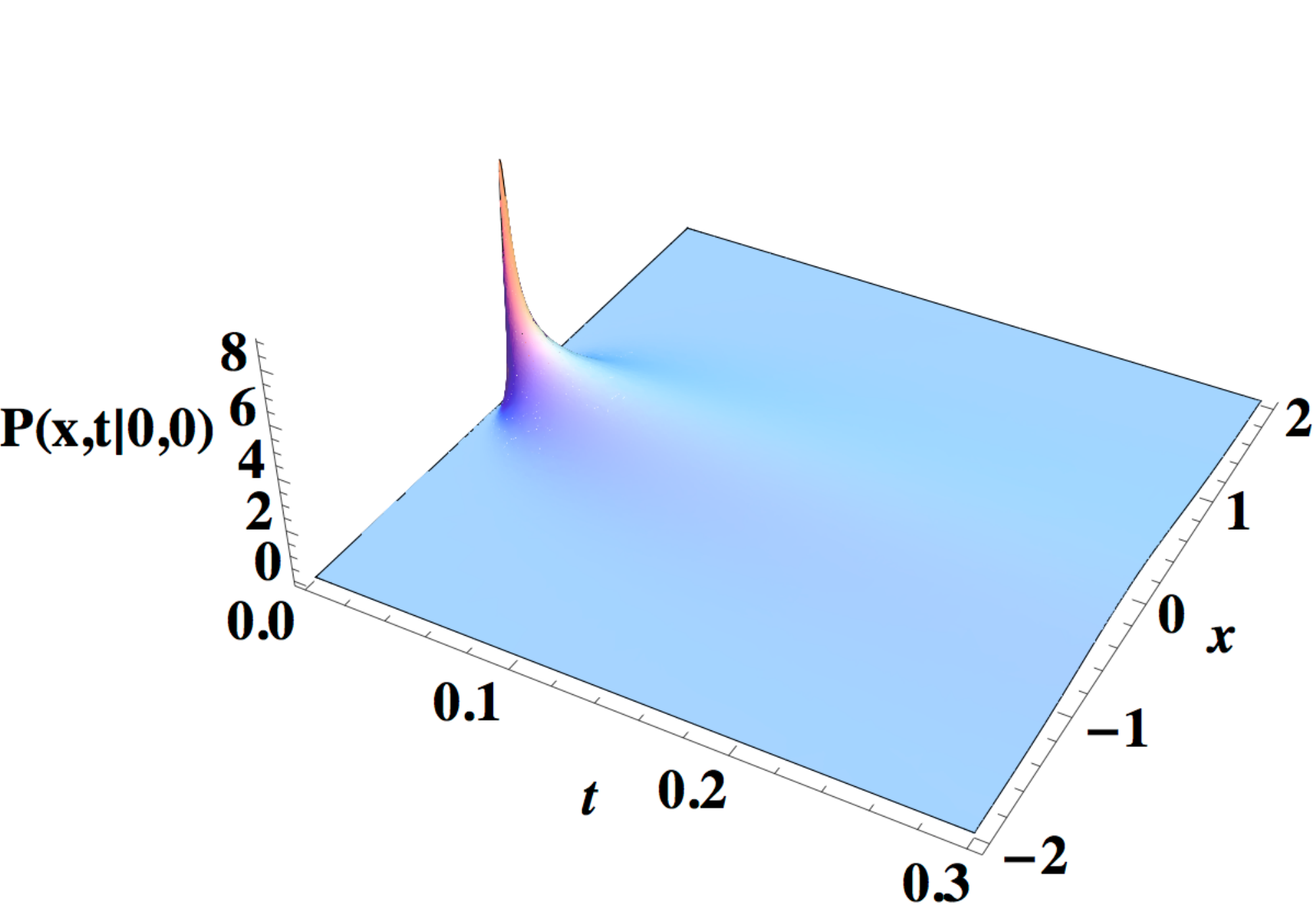}
  \caption{The probability of being in position $x$ at a time $t$ differs
  from the equilibrium probability distribution $\rho_{eq}(x)$,
  if we know the position of the particle at a previous time.
  }
  \label{fig1C2}
\end{subfigure}
\caption{A particle diffusing according to $\dot{x} = -x + \eta(t)$ with
  diffusion coefficient $D=1$ moves as in Figure~\ref{fig1A2}. Over infinite
  time, the particle experiences positions distributed according to the
  probability density function in Eqn. \ref{eq:StationaryDist}; see
  Figure~\ref{fig1B2}. If the previous particle position is known, a future
  particle position can be determined with less uncertainty than if no previous
  particle position is known, as shown in Figure~\ref{fig1C2}.
  }
\label{fig:OneDLangevin}
\end{figure}

Consider an $n$-dimensional nonlinear Langevin equation:
\begin{align*}
\frac{dx}{dt} = - D \nabla U(x) + \eta(t) ~,
\end{align*}
where $x \in \mathbb{R}^n$, $U(x)$ is an analytic potential function and $\eta(t)$ is zero-mean white noise with diffusion matrix $D$:
$\langle \eta_i(t)\rangle = 0$ and
$\langle \eta_i(t)\eta_j(t') \rangle = D_{ij} \delta(t-t')$.
The diffusion coefficients $D_{ij}=D_{ji}$ are assumed to be independent of $x$
and such that $\det D \neq 0$.  The following (well-known) stationary
distribution is derived by converting the stochastic differential equation into
its Fokker-Planck equation form:
\begin{align}
\rho_{eq}(x) = \frac{1}{Z} \exp\left(-2U(x)\right) ~,
  \label{eq:StationaryDist}
\end{align}
where $Z = \int e^{-2U(x)} dx$.
We assume that this is the stationary probability distribution experienced by
the particle and that it is normalizable: $Z<\infty$.
(See Fig. \ref{fig:OneDLangevin} for simulation results in one dimension.)

The time-discretization normalized entropy of a measurement is:
\begin{align}
H_0 & = - \frac{1}{\tau} \int\rho_{eq}(x)\log\rho_{eq}(x) dx ~.
\label{eq:H0}
\end{align}
The conditional entropies $H[X_{\tau}|X_0]$ and $H[X_{\tau}|X_{-\tau}]$ in
Eqns.~\ref{eq:1main}-\ref{eq:2main} simplify if the conditional probabilities
$\Prob(X_{\tau}|X_0)$ and $\Prob(X_{\tau}|X_{-\tau})$ are Gaussians, since:
\begin{align*}
H[X_{\tau}|X_0] &= \int \rho_{eq}(x') H[X_{\tau}|X_0 = x'] dx' \\
H[X_{\tau}|X_{-\tau}] &= \int \rho_{eq}(x') H[X_{\tau}|X_{-\tau}=x'] dx'
\end{align*}
and
\begin{align}
H[X_{\tau}|X_{0}=x'] &= \frac{1}{2}
  \log (2\pi e |\det \text{Var}(X_{\tau})_{p(X_{\tau}|X_{0}=x')}|)
  \label{eq:condentropy1} \\
H[X_{\tau}|X_{-\tau}=x'] &= \frac{1}{2}
  \log (2\pi e |\det \text{Var}(X_{\tau})_{p(X_{\tau}|X_{-\tau}=x')}|) ~.
\label{eq:condentropy2}
\end{align}
Appendix \ref{sec:Greens} shows that the conditional distributions
$\Prob(X_{\tau}|X_0)$ and $\Prob(X_{\tau}|X_{-\tau})$ are Gaussian to
$o(\tau)$ over a region of $\mathbb{R}^n$ with measure arbitrarily close
to $1$. The entropies of these Gaussians are calculable to leading and subleading order in $\tau$ using a linearized version of the nonlinear Langevin equation about the initial position:
\begin{align*}
\frac{dx}{dt} = \nabla U(x)|_{x=x'} + A(x') (x-x') + \eta(t)
  + O(||x-x'||^2) ~,
\end{align*}
where $A(x')$ is a matrix with entries
$\left(A(x')\right)_{ij} = \partial (D\nabla U)_j / \partial x_i$.
(This is similar but not identical to the approximation used in
\cite{Mart13a}. Appendix \ref{sec:TiPi} comments on the differences.)
From Appendix \ref{sec:Greens}, we have that:
\begin{align}
\text{Var}(X_{\tau})_{p(X_{\tau}|X_{0}=x')} &=
  D\tau + \frac{\nabla\mu(x) D + D (\nabla\mu(x))^{\top}}{2} \tau^2
  + O(\tau^3)
\label{eq:var1}
\end{align}
and, similarly,
\begin{align}
\text{Var}(X_{\tau})_{p(X_{\tau}|X_{-\tau}=x')} &=
  2D\tau + 2(\nabla\mu(x) D + D (\nabla\mu(x))^{\top}) \tau^2  + O(\tau^3) ~. \label{eq:var2}
\end{align}
Substituting Eqns.~\ref{eq:var1} and \ref{eq:var2} into
Eqns.~\ref{eq:condentropy1} and \ref{eq:condentropy2}, respectively, gives
with some algebra:
\begin{align}
H[X_{\tau}|X_{-\tau}] &= -\tau \int \rho_{eq}(x)
  \nabla\cdot( D\nabla U(x)) dx
  + \log\sqrt{2^{n+1} \pi e |\det D|\tau^n}
\label{eq:condentropyfinal1}
\end{align}
and
\begin{align}
H[X_{\tau}|X_{0}] &= - \frac{\tau}{2} \int \rho_{eq}(x)
  \nabla\cdot (D\nabla U(x)) dx
  + \log\sqrt{2\pi e |\det D|\tau^{n}} ~.
\label{eq:condentropyfinal2}
\end{align}
Substituting Eqn.~\ref{eq:condentropyfinal2} into Eqn.~\ref{eq:1main}, we
find that:
\begin{align}
\hmu (\tau) &= \frac{n\log\sqrt{2\tau}}{\tau}
  + \frac{\log\sqrt{\pi e |\det D|}}{\tau}
  - \frac{1}{2}\int \rho_{eq}(x) \nabla \cdot (D\nabla U) dx
  + o(1) ~.
\label{eq:hmu2}
\end{align}
The leading order term is recognizable as an $(\epsilon,\tau)$-entropy rate of
the Ornstein-Uhlenbeck process \cite{Gasp93a}, except that the $\epsilon$ has
been regularized away since we used Shannon's differential entropy.
Substituting Eqns.~\ref{eq:condentropyfinal1} and \ref{eq:condentropyfinal2}
into Eqn.~\ref{eq:2main}, we find the bound information rate:
\begin{eqnarray}
\bmu (\tau) &=& \frac{n \log \sqrt{2}}{\tau} - \frac{1}{2}\int \rho_{eq}(x)
\nabla\cdot (D \nabla U(x)) dx + o(1) ~.
  \label{eq:bmu2}
\end{eqnarray}
Thus, the rate of active information storage depends on the dimension of the state space to leading order in $\tau$, but its nondivergent part depends on the average curvature of the potential.
%If the drift term is curl free and $\mu = -\nabla U$, then $\text{tr}(\nabla \mu) = -\nabla^2 U$.
%In the limit that the diffusion coefficient vanishes, $D\rightarrow 0$, the bound information rate will pick up the curvature of the potential $U(x)$ at the potential's global minimum, a point that we explore later.

From these quantities all other anatomy measures follow. Substituting Eqns.
\ref{eq:hmu2} and \ref{eq:bmu2} into Eqn.~\ref{eq:rmu}, we find that the
ephemeral information is:
\begin{align}
\rmu (\tau) &= \frac{n\log\sqrt{\tau}}{\tau}
  + \frac{\log\sqrt{2 \pi e |\det D|}}{\tau} + o(1) ~.
\label{eq:rmu2}
\end{align}
Unsurprisingly, the dissipated information---that entropy created in the
present useful for neither predicting nor retrodicting---depends only on the
noisiness of the dynamics and not the drift.

\vspace{0.2in}
\begin{minipage}{0.95\linewidth}
\begin{center}
\begin{tabular}{C{1.35in} C{0.65in} C{0.5in} C{1.4in} C{2.0in}}
\toprule[1.5pt]
\bf Information rates & \bf Definition & \multicolumn{3}{c}{\bf Terms} \\
& & $O(\tau^{-1} \log\tau)$ & $O(\tau^{-1})$ & $O(1)$ \\
\midrule
Stored $H_0 = \Cmu(\tau)$ & $\frac{H[X_0]}{\tau}$ & $0$ & $-\int \rho_{eq}(x) \log \rho_{eq}(x)dx$ & $0$ \\
\hline
$\tau$-Entropy $\hmu (\tau)$ & $\frac{H[X_0|X_{:0}]}{\tau}$ &
$\frac{n}{2}$ & $\log\sqrt{2\pi e |\det D|} +n\log\sqrt{2}$ &
$-\tfrac{1}{2} \int \nabla \cdot (D\nabla U(x)) \rho_{eq}(x) dx$ \\
\hline
Bound $\bmu (\tau)$ & $\frac{I[X_0;X_{\tau:}|X_{:0}]}{\tau}$ & $0$
& $n\log\sqrt{2}$ & $-\frac{1}{2} \int \nabla \cdot (D\nabla U(x)) \rho_{eq}(x) dx$ \\
\hline
Ephemeral $\rmu (\tau)$ & $\frac{H[X_0|X_{:0},X_{\tau:}]}{\tau}$ & $\frac{n}{2}$ & $\log\sqrt{2\pi e |\det D|}$ & $0$ \\
\hline
Enigmatic $\qmu (\tau)$ & $\frac{I[X_{:0};X_0;X_{\tau:}]}{\tau}$ & $-\frac{n}{2}$ & $-\int \rho_{eq}(x) \log \rho_{eq}(x)dx-n\log 2-\log\sqrt{2\pi e|\det D|}$ & $\int \nabla\cdot (D\nabla U(x)) \rho_{eq}(x) dx$ \\
\hline
Elusive $\sigmamu (\tau)$ & $\frac{I[X_{:0};X_{\tau:}|X_0]}{\tau}$ & $0$ & $0$ & $0$ \\
\bottomrule[1.25pt]
\end {tabular}\par
\bigskip
\captionof{table}{Information anatomy of first-order, $n$-dimensional nonlinear
  Langevin dynamics: $\dot{x} = -D\nabla U(x)+ \eta(t)$, where $U(x)$ is
  analytic in $x$ and $\eta(t)$ is zero-mean white noise with invertible
  diffusion matrix $D$, $\langle \eta(t) \eta(t')^{\top} \rangle =
  D\delta(t-t')$. Stationary distribution $\rho_{eq}(x)\propto \exp(-U(x))$
  assumed normalizable.
  }
\label{tab:title} 
\end{center}
\end{minipage}

Finally, the enigmatic information---that shared between past, future, and
present---follows by substituting Eqns.~\ref{eq:H0}-\ref{eq:bmu2} into
Eqn.~\ref{eq:qmu}:
\begin{align*}
\qmu (\tau) = -\frac{1}{\tau}\int \rho_{eq}(x) \log \rho_{eq}(x) dx
  & - \frac{n\log(2\sqrt{\tau})}{\tau}
  - \frac{\log\sqrt{\pi e |\det D|}}{\tau} \nonumber \\
  & - \int \rho_{eq}(x) \nabla\cdot(D\nabla U)  dx
  + O(\tau) ~.
\end{align*}
It is interesting to consider how $\qmu$ changes as the stochasticity of
the system increases: The stationary distribution $\rho_{eq}(x)$ flattens out,
leading to an unbounded increase in $H_0$. This is counteracted by
an unbounded increase in the entropy rate.

We can also bound the bound information rate when $D$ is positive semidefinite and
$\nabla U$ grows more slowly than $e^{-2U}$ with $||x||$. Then, integration
by parts applied to Eqn.~\ref{eq:bmu2} gives;
\begin{align*}
\bmu (\tau) &= \frac{n\log\sqrt{2}}{\tau}
  - \frac{1}{2} \int (\nabla U)^{\top}
  D (\nabla U)~\frac{e^{-U(x)}}{Z} dx ~.
% \label{eq:bmu3}
\end{align*}
When $D$ is positive semidefinite, $v^{\top}Dv \geq 0$ for any vector $v$,
then:
\begin{align*}
\bmu (\tau) \leq \frac{n\log\sqrt{2}}{\tau} ~.
\end{align*}
Therefore, $\bmu (\tau)$ is maximized for a positive semidefinite diffusion matrix when the potential well is as flat as possible, while maintaining
$Z < \infty$.

%In the deterministic limit, when $D=DI_{n\times n}$ and $D\rightarrow 0$, the stationary probability distribution tends again to delta distributions at the global minima $\{x^*_1,...,x^*_m\}$, defined by $U(x^*_i) = \min_x U(x)~\forall~i=1,...,m$.  In this limit, from Eqn.~\ref{eq:bmu2},
%\begin{eqnarray}
%\lim_{D\rightarrow 0} b_{\mu}(\tau) &=& \frac{n\log\sqrt{2}}{\tau} -\frac{1}{2m} \sum_{i=0}^{m} \nabla^2 U(x)|_{x=x^*_i}.
%\end{eqnarray}
%If $\theta$ parameterizes the potential $U(x)$, and that $U(x)$ is a continuous function of $\theta$, then it is clear that $b_{\mu}(\tau)$ would also be a continuous function of the parameters $\theta$.  This continuity is apparent in the cusp catastrophe example in Section \ref{sec:cusp_catastrophe}.

\subsection{Linear Langevin equation with Noninvertible Diffusion}
\label{sec:first-orderlinearLangevin}

What if the invertibility of the diffusion matrix is relaxed? In particular,
do we still have qualitatively the same information anatomy if a subsystem of
the stochastic dynamical system evolves deterministically? How does this affect the information generation and storage properties? To this end, suppose
$x = \left( x_d \quad x_n \right)^{\top}$ with $x \in \mathbb{R}^k$
and $m = \dim (x_d)$, where $x_d$ evolves deterministically and $x_n$
stochastically:
\begin{align}
\frac{dx_d}{dt} &= A_d + B_{dd} x_d + B_{dn} x_n \label{eq:linLang1} \\
\frac{dx_n}{dt} &= A_n + B_{nd} x_d + B_{nn} x_n + \eta(t)
  ~.
\label{eq:linLang2}
\end{align}
Again, $\eta(t)$ is white noise with $\langle \eta(t)\rangle = 0$ and
$\langle \eta(t)\eta(t')^{\top}\rangle = D\delta(t-t')$, where $D$ is
invertible. Taken together, though, this is a linear Langevin equation for
$x$ with a noninvertible diffusion matrix. Naively assuming that the
deterministic subsystem evolves with a small amount of noise,
Eqn.~\ref{eq:bmu2} would apply and give, for example, to $O(\tau)$:
\begin{align*}
\bmu = \frac{n \log 2}{2 \tau} +
\frac{\text{tr}(B_{dd})+\text{tr}(B_{nn})}{2} ~.
\end{align*}
But this assumption would be incorrect; the noiseless limit is singular.
%Let $\Sigma$ be the variance of $x$ in the stationary probability distribution $\rho_{eq}(x)$, which is a Gaussian since this is essentially a linear dynamical system with Gaussian noise.  Then
%\begin{equation}
%H_0 = \frac{1}{\tau} \log \sqrt{2\pi e|\det \Sigma|}.
%\end{equation}

Since Eqns.~\ref{eq:linLang1} and \ref{eq:linLang2} specify a linear Langevin equation for $x$, its Green's function is Gaussian.
%In Appendix \ref{sec:linearLangevin}, assuming that $B_{dn} B_{dn}^{\top}$ is invertible, we derive the conditional entropy to $O(\tau^2)$ to be
%\begin{eqnarray}
%H[X_{\tau}|X_0] &=& \log\sqrt{2\pi e|\det D_{nn}| |\det B_{dn} D_{nn} B_{dn}^{\top}|} + n\log\frac{\tau}{12^{1/4}} + \frac{\text{tr}(B_{dd})-3\text{tr}(B_{nn})}{2} \tau.
%\end{eqnarray}
From App.~\ref{sec:linearLangevin}, to $O(\tau)$ the entropy rate is:
\begin{align*}
\hmu (\tau) = \frac{(n + 2m) \log \tau}{2 \tau}
  & - \frac{m\log\sqrt{12}}{2\tau}
  + \frac{\log \sqrt{2\pi e |\det D_{nn}| |\det B_{dn} D_{nn}
  B_{dn}^{\top}|}}{\tau} \nonumber \\
  & + \frac{\text{tr}(B_{dd})-3\text{tr}(B_{nn})}{2}
\end{align*}
and the bound information is:
\begin{align}
\bmu (\tau) &= \frac{(n + 2m) \log 2}{2 \tau}
  + \frac{\text{tr}(B_{dd})-3\text{tr}(B_{nn})}{2}
  ~.
\label{eq:bmu_linear}
\end{align}
Applying Eqn.~\ref{eq:rmu}, the ephemeral information rate is to $O(\tau)$:
\begin{align}
\rmu (\tau) &= \frac{n+2m}{2} \frac{\log (\tau/2)}{\tau}
  -\frac{m\log\sqrt{12}}{2\tau}
  + \frac{\log \sqrt{2\pi e |\det D_{nn}| |\det B_{dn} D_{nn} B_{dn}^{\top}|}}{\tau}
  ~.
\label{eq:rmu_linear}
\end{align}

These answers are very different from those derived assuming that $x$'s
deterministic subsystem $x_d$ evolves with an infinitesimal amount of noise.
The bound information in Eqn.~\ref{eq:bmu_linear} differs from that found from
naive application of Eqn.~\ref{eq:bmu2} in two ways. First, the pre-factor for
the $\log 2 / \tau$ divergence is $n/2 + m$ rather than $n/2$. That is, the
difference counts the dimension $m$ of the deterministically evolving state
space $x_d$. Thus, the deterministic subsystem allows for the active storage of
more of the spontaneously generated stochasticity.  Second, $\bmu$'s $O(1)$ term involves $\text{tr}(B_{dd})-3\text{tr}(B_{nn})$ rather than
$\text{tr}(B_{dd})+\text{tr}(B_{nn})$.

The ephemeral information in Eqn.~\ref{eq:rmu_linear} differs from a naive
application of Eqn.~\ref{eq:rmu2} in two new ways. First, the expression in
Eqn.~\ref{eq:rmu_linear} has an additional $O(1/\tau)$ factor that is
linearly proportional to the dimension $m$ of the deterministic subsystem. And, second, the term $\log (2\pi e |\det D_{nn}| |\det B_{dn} D_{nn} B_{dn}^{\top}|)$ can be interpreted by supposing that $B_{dn} D_{nn} B_{dn}^{\top}$ is the effective diffusion matrix felt by the deterministically evolving states.

These information anatomy quantities are therefore sensitive to the process's
underlying noise architecture.

\section{Examples}

To illustrate how the information measures are helpful and interesting
summaries of nonlinear Langevin dynamics, let's consider several examples.

\subsection{Stochastic Gradient Descent in One Dimension}
\label{sec:cusp_catastrophe}

Consider a first-order nonlinear Langevin dynamics for $x \in \mathbb{R}$
in which:
\begin{align*}
\frac{dx}{dt} = -\frac{dU(x)}{dx} + \eta(t) ~,
\end{align*}
where $\langle\eta(t)\rangle=0$ and
$\langle\eta(t)\eta(t')\rangle=2D\delta(t-t')$. The stationary distribution
is:
\begin{align*}
\rho_{eq}(x) = \frac{1}{Z} e^{-U(x)/D} ~,
\end{align*}
with $Z$ a normalization factor:
\begin{align*}
Z = \int_{-\infty}^{\infty} e^{-U(x)/D} dx ~.
\end{align*}
We require that $Z < \infty$.

This process's elusive information is zero and the ephemeral information rate
is the strength of the noise. But the bound information is:
\begin{equation}
\bmu (\tau) = \frac{\log\sqrt{2}}{\tau}
  - \frac{1}{Z} \int_{-\infty}^{\infty} e^{-U(x)/D} \frac{d^2 U(x)}{d x^2} dx
  ~.
\label{eq:bmu_stochastic}
\end{equation}
Using integration by parts, this can be rewritten:
\begin{align*}
\bmu (\tau) = \frac{\log\sqrt{2}}{\tau}
  - \frac{1}{D} \int_{-\infty}^{\infty}
  \left(\frac{dU}{dx}\right)^2 \frac{e^{-U(x)/D}}{Z} dx ~.
\end{align*}
So, $\bmu$ is sensitive to the average curvature of the potential or,
equivalently, to the average squared drift normalized by the diffusion constant.

In the deterministic limit, this expression simplifies.
Suppose that $\{x^*_1,...,x^*_m\}$ are the global minima of the potential
function: $U(x^*_i) = \min_x U(x)$, for $i=1,...,m$.
It follows that $\lim_{D\rightarrow 0} \frac{e^{-U(x)/D}}{Z} = \sum_{i=1}^m
\frac{1}{m}\delta(x-x^*_i)$.  Applying this limit to
Eqn.~\ref{eq:bmu_stochastic}, we have:
\begin{align*}
\lim_{D\rightarrow 0} b_{\mu}(\tau) = \frac{\log\sqrt{2}}{\tau} - \frac{1}{2m}\sum_{i=1}^m \frac{d^2 U(x)}{dx^2}|_{x=x^*_i}
  ~.
\end{align*}
This limit is a little strange. If $D=0$ exactly, so that we have
deterministic gradient descent, then the stationary time series consists of a
single measurement. The information anatomy becomes rather trivial. There is
no uncertainty in the present measurement and the past, present, and future
share no information. If $D$ is nonzero, no matter how small, however, then there is finite uncertainty in a measurement and the past, present, and
future share information with one another.

\begin{figure}
\checkoddpage
  \edef\side{\ifoddpage l\else r\fi}%
  \makebox[\textwidth][\side]{
\begin{subfigure}[h]{.5\linewidth}
	\centering
\includegraphics[width=\textwidth]{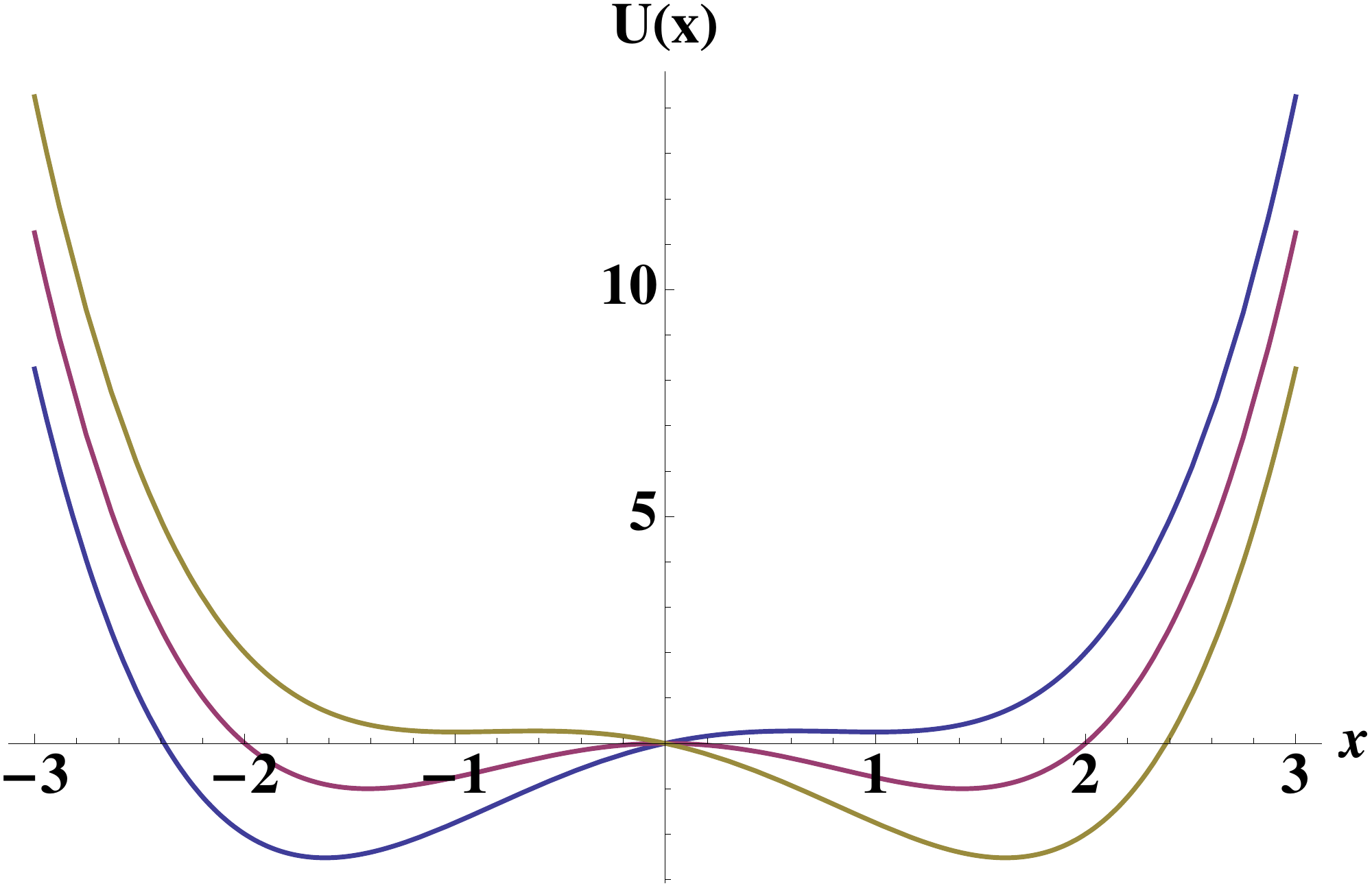}
\caption{Example potentials $U(x)$ for various $r$ and $h$: Blue,
  $r=2$ and $h=-1$; purple, $r=2$ and $h=0$; and yellow, $r=2$ and $h=1$.
  }
\label{fig2a}
\end{subfigure}

\begin{subfigure}[h]{.5\linewidth}
\centering
\includegraphics[width=0.8\textwidth]{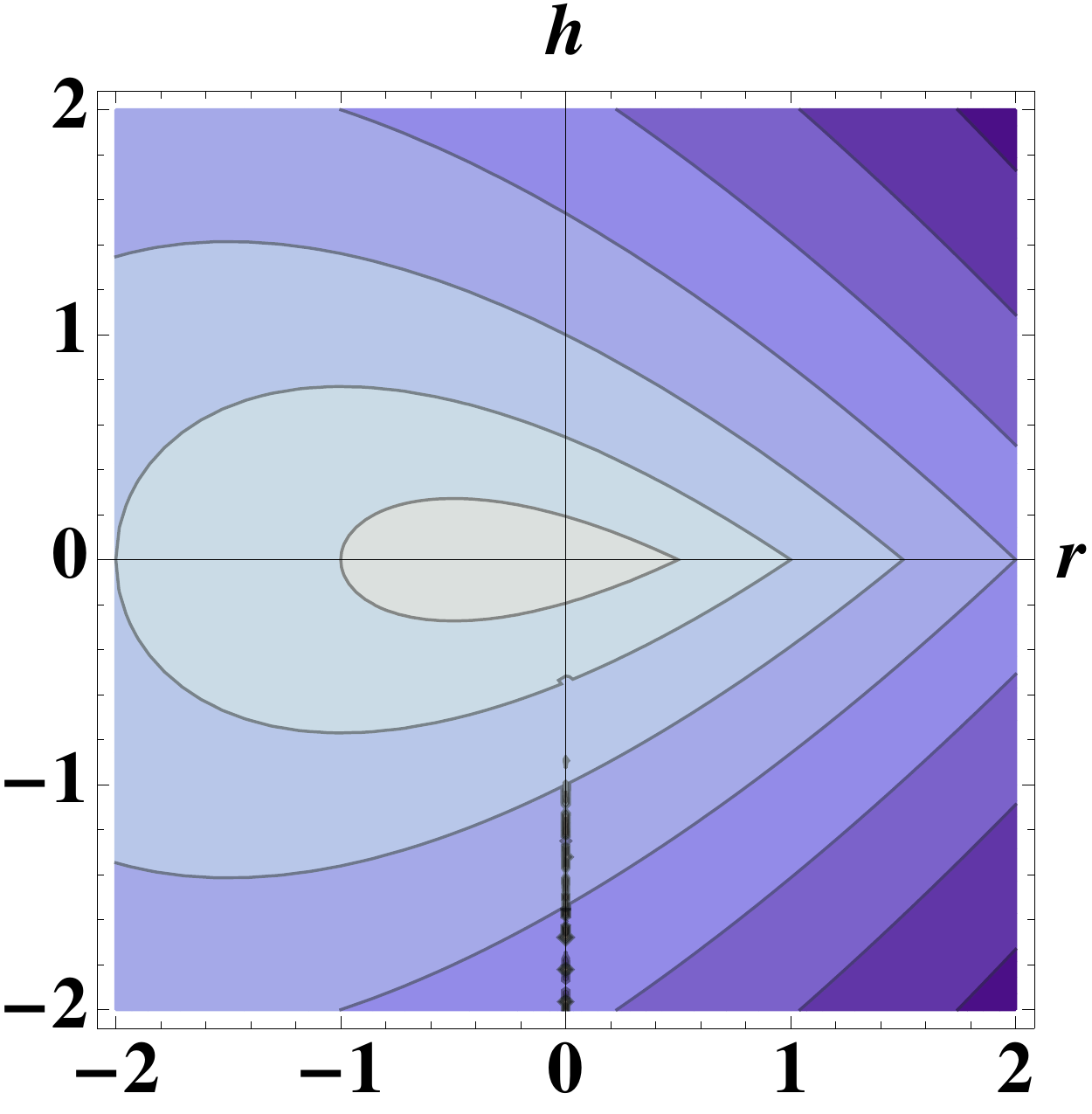}
	\caption{$\lim_{D\rightarrow 0} b_{\mu}(\tau) - \tau^{-1} \log\sqrt{2}$
	as a function of $r$ and $h$: $b_{\mu}(\tau)$ is nondifferentiable along
	$h=0$, $r\geq 0$ with respect to $h$.}
	\label{fig2B}
\end{subfigure}
}

\checkoddpage
  \edef\side{\ifoddpage l\else r\fi}%
  \makebox[\textwidth][\side]{
\begin{subfigure}[h]{.5\linewidth}
	\centering
\includegraphics[width=\textwidth]{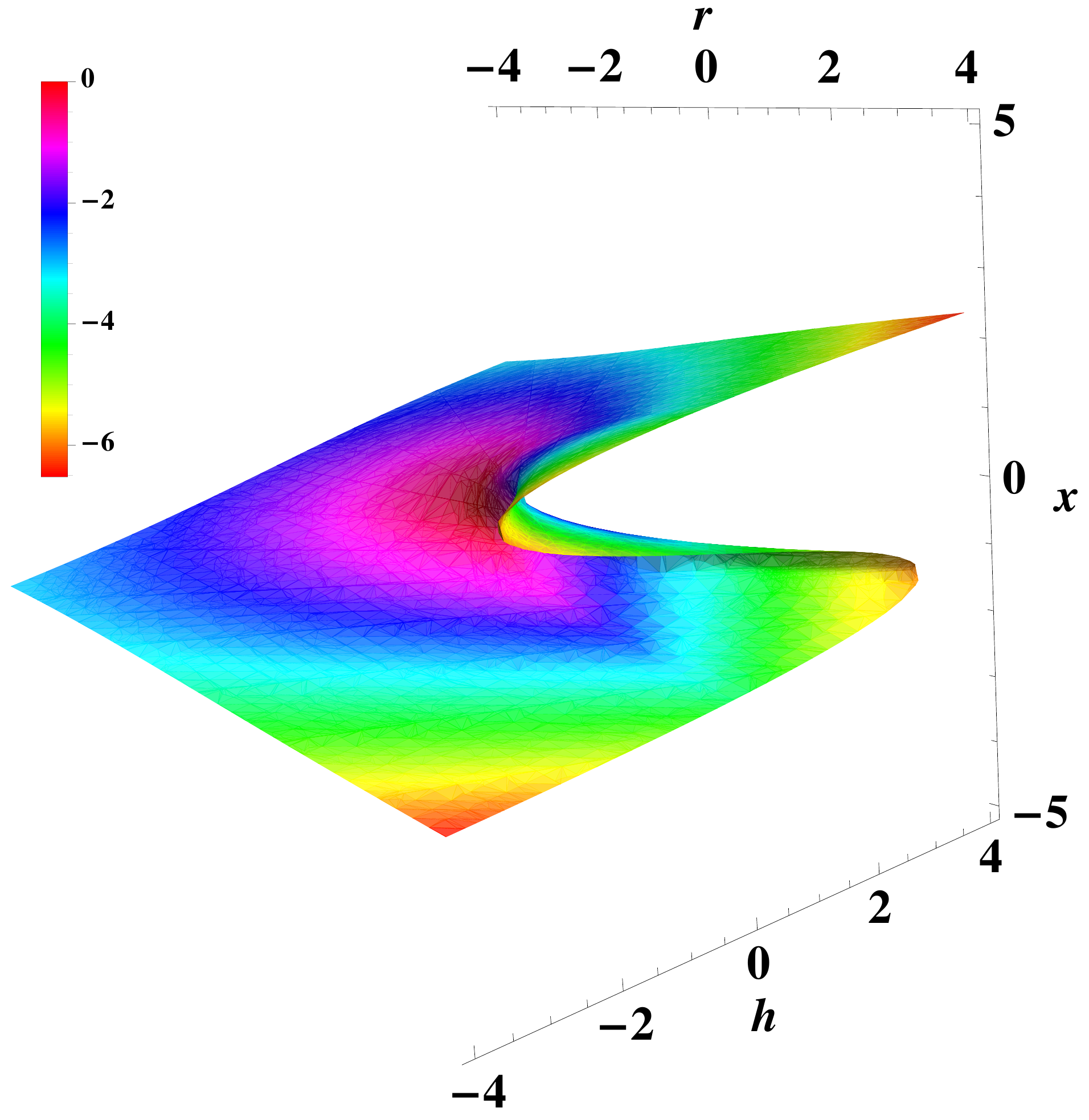}
\caption{Bound information on the cusp catastrophe equilibrium surface:
  Height gives the fixed points as a function of $r$ and $h$. Color hue is
  proportional to the deterministic limit
  $\lim_{D\rightarrow 0} \bmu (\tau) - \frac{\log\sqrt{2}}{\tau}$
  at each $r$ and $h$.
  }
\label{fig2A}
\end{subfigure}

\begin{subfigure}[h]{.5\linewidth}
\centering
\includegraphics[width=1.0\textwidth]{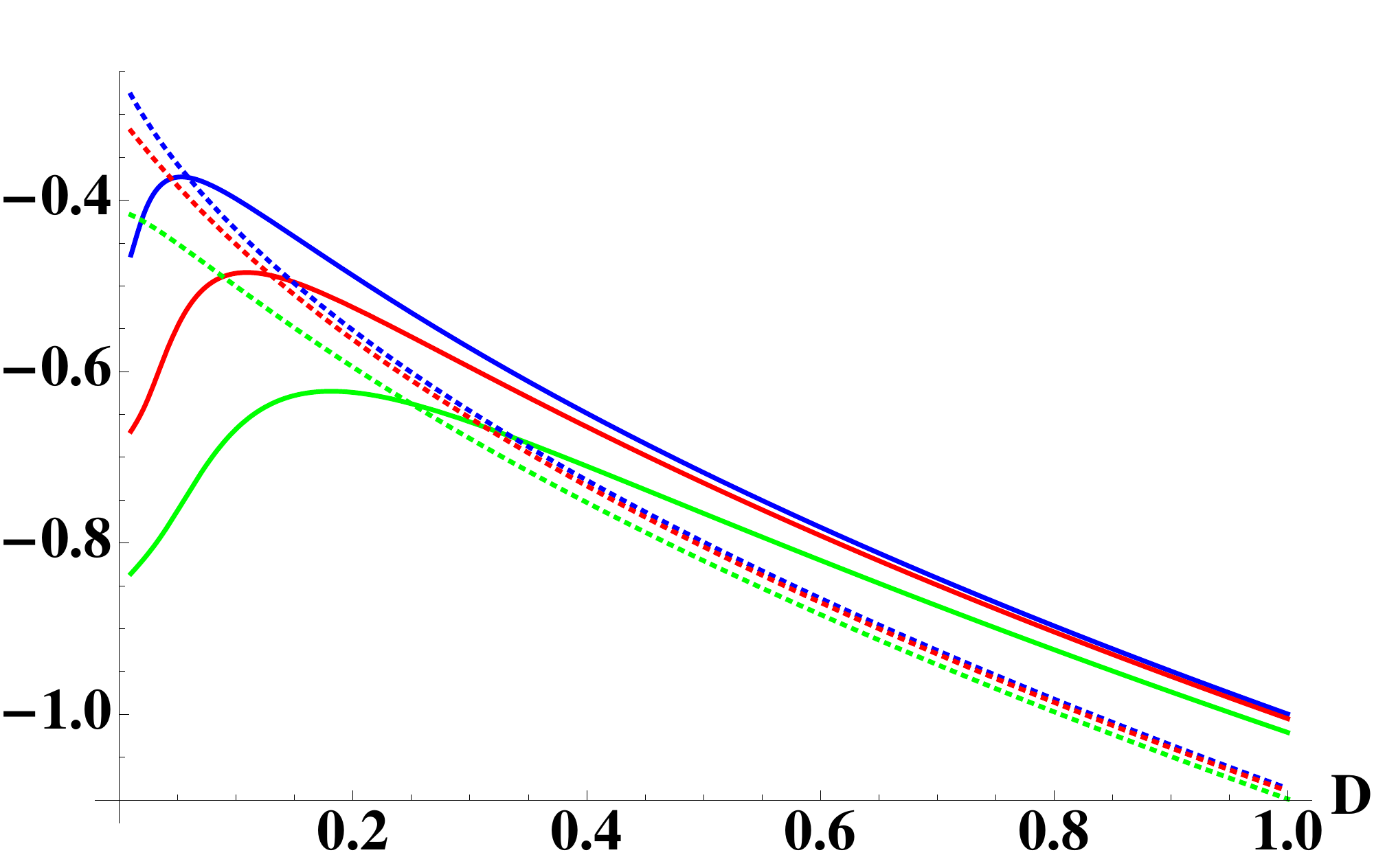}
	\caption{$b_{\mu}(\tau) - \tau^{-1} \log\sqrt{2}$ as a function of $D$ for several values of $r$ and $h$: dotted lines are $r=-0.5$ and solid lines are $r=0.5$; blue is $h=0$, red is $h=0.1$, and green is $h=0.2$.}
	\label{fig2C}
\end{subfigure}
}
\caption{Information anatomy of the stochastic cusp catastrophe: (a) Shifting
  from a double-well to a single-well potential as $r$ and $h$ are varied.
  (b) Contour plot of the system-dependent part of the bound information rate
  $\bmu (\tau)$ as a function of $r$ and $h$, highlighting the global minimum
  $x^*$ changing discontinuously as $h$ moves through zero. (c) $\bmu (\tau)$
  as it varies over the cusp equilibria surface. (d) Bound information rate
  is maximized at nonzero stochasticity $D$ for double-well, but not
  single-well potentials.
  }
\label{fig2}
\end{figure}

As a concrete example, consider the canonical form for the cusp catastrophe
\cite{Post78a}:
\begin{align*}
\frac{dx}{dt} = h+rx-x^3 + \eta(t) ~,
\end{align*}
with additive noise where $\langle \eta(t)\eta(t')\rangle = 2D\delta(t-t')$.
The potential function is $U(x) = \frac{1}{4} x^4 - \frac{1}{2}rx^2 - hx$,
and the corresponding bound information in the noiseless limit is:
\begin{align*}
\lim_{D\rightarrow 0} \bmu (\tau,r,h)
  = \frac{\log\sqrt{2}}{\tau} + \frac{r-3 (x^*(r,h))^2}{2} ~.
\end{align*}
The global minimum $x^*(r,h)$ is \textit{not} everywhere differentiable in $r$
and $h$, and this appears also in $\bmu (\tau,r,h)$. See Figure \ref{fig2}. The
contour of nondifferentiability is $h = 0$ for $r > 0$. Along the contour,
the potential is symmetric, there are suddenly two global minima of $U(x)$
with $x^*_1 = -x^*_2$ and so the sign of $x^*$ changes discontinuously across
$h = 0$.

%\begin{figure}
%\centering
%\includegraphics[width=0.45\textwidth]{bmu_cusp_catastrophe}
%\caption{Plotted is $\lim_{D\rightarrow 0} b_{\mu}(\tau) - \frac{\log\sqrt{2}}{\tau}$ as a function of $r$ and $h$.}
%\label{fig2}
%\end{figure}

%\begin{subfigure}[h]{.5\linewidth}
%	\centering
%\includegraphics[width=\textwidth]{bmu_cusp_catastrophe4.pdf}
%\caption{Global minimum $x^*$ of $U(x)$ as a function of $r$ and $h$.
%  At $h=0$, there are two global minima symmetric about the origin.
%  }
%\label{fig2b}
%\end{subfigure}
%}

Interestingly, for double-well potentials but not single-well potentials,
$b_{\mu}(\tau)$ is maximized at a nonzero noise level $D>0$.  At some level,
this is completely counterintuitive.  Adding noise only decreases the
predictability of a process. However, adding noise in the present affects the
future in a way that cannot be predicted from the past.  Since $b_{\mu}(\tau)$
measures the amount of information shared between the present and future which
is not shared with the past, there is (for some double-well potentials) a level
of stochasticity that maximizes $b_{\mu}(\tau)$. This is shown in Figure \ref{fig2C}.

\subsection{Particles Diffusing in a Heat Bath}
\label{sec:particles_heatbath}

Suppose $N$ particles with positions $x_1,...,x_N$ and masses $m_1,...,m_N$
diffuse according to the potential function $U(x_1,...,x_N)$ in a heat bath
of temperature $T$. Let $\textbf{x}$ denote the vector of concatenated
particle positions. When the inertial terms $m_i d^2 x_i/ dt^2$ are
negligible, an overdamped Langevin equation can be used to approximate the particles' trajectories:
\begin{align*}
\frac{d\textbf{x}}{dt}
  & = \frac{1}{\gamma} M^{-1} \nabla U(\textbf{x}) + \eta(t) \\
\langle \eta_i(t)\rangle & = 0 \\
\langle \eta_i(t)\eta_j(t') \rangle
  & = \frac{2k_B T}{\gamma m_i} \delta_{i,j} \delta(t-t')
  ~ .
\end{align*}
$M$ is a diagonal matrix whose entries are the particle masses and the
parameter $\gamma$ is a friction coefficient that controls how strongly the particles couple to the heat bath. The stationary distribution of positions $x$ is the Boltzmann distribution:
\begin{align*}
\rho_{eq}(\textbf{x})
 = \frac{1}{Z} \exp\left(-\frac{U(\textbf{x})}{k_B T}\right) ~,
\end{align*}
where $Z$ is the partition function:
\begin{align*}
Z = \int \exp\left(-\frac{U(\textbf{x})}{k_B T}\right) d\textbf{x} ~.
\end{align*}
From Eqn.~\ref{eq:H0}, the normalized single-measurement entropy is:
\begin{align*}
H_0 = \frac{1}{\tau}
  \left(\frac{\langle U(\textbf{x})\rangle}{k_B T} + \ln Z \right) ~.
\end{align*}
where
\begin{align*}
\langle U(\textbf{x})\rangle
  = \int U(\textbf{x}) \frac{e^{-U(\textbf{x})/k_B T}}{Z} d\textbf{x} ~,
\end{align*}
which is simply proportional to the familiar definition of entropy in physics.

For notational ease, let $\bar{m}$ denote the geometric mean of the masses:
\begin{align*}
\bar{m} = \left(\prod_{i=1}^N m_i\right)^{1/N} ~,
\end{align*}
$k_i$ the effective ``spring constant'' for the $i^{th}$ particle:
\begin{align*}
k_i = \int \frac{\partial ^2
  U(\textbf{x})}{\partial x_i^2}
  \frac{e^{-U(\textbf{x})/k_B T}}{Z} d\textbf{x} ~,
\end{align*}
and $\omega_i$ the effective ``oscillation frequency'' for the $i^{th}$ 
particle:
\begin{align*}
\omega_i = \sqrt{k_i/m_i} ~.
\end{align*}
From Eqn.~\ref{eq:hmu2}, the entropy rate is:
\begin{align*}
\hmu (\tau) &=
  \frac{N\log\sqrt{\frac{4 k_B T \tau}{\gamma\bar{m}}}}{\tau}
  + \frac{\log\sqrt{\pi e}}{\tau}
  - \frac{1}{2\gamma} \sum_{i=1}^N \omega_i^2+o(1)
  ~.
\end{align*}
From Eqn.~\ref{eq:bmu2}, the bound information is to similar order:
\begin{align*}
\bmu (\tau) &=
  \frac{N\log\sqrt{2}}{\tau} - \frac{1}{2\gamma} \sum_{i=1}^N \omega_i^2+o(1)
  ~.
\end{align*}
From Eqn.~\ref{eq:rmu2}, the ephemeral information rate is:
\begin{align*}
\rmu (\tau) &=
  \frac{N\log\sqrt{\frac{2k_B T \tau}{\gamma \bar{m}}}}{\tau}
  + \frac{\log\sqrt{2\pi e}}{\tau} + o(1)
  ~.
\end{align*}

Several information measures appear dimensionally incorrect. This is a
perennial concern when calculating the differential entropy of random
variables that themselves have units. The probability density over those
variables also has a dimension and this leads to differential entropies that
involve the log of a number with dimension. Implicitly, however, we chose
a standard unit system such that all quantities are dimensionless.

%In any case, it seems that everything except for the time-normalized measurement entropy contains very little information about the underlying potential function and has more to do with the particles' coupling to the heat bath.

All of these quantities are extensive in $N$. The normalized entropy per
measurement $H_0$ is proportional to the Boltzmann entropy by a factor of
$k_B / \tau$. The entropy rate $\hmu (\tau)$ and ephemeral information
$\rmu (\tau)$ increase logarithmically with the mean squared velocity
$\sqrt{\langle v^2\rangle} = \frac{k_B T}{m}$. The bound information
$\bmu (\tau)$ increases when there is a larger $\gamma$; that is, when
there is stronger coupling between the particles and the heat bath or when there is a
smaller average oscillation frequency $\sum_{i=1}^N \omega_i^2$. Since $\gamma
\geq 0$ and $\omega_i^2 \geq 0$, the bound information is bounded above by
$\bmu (\tau) \leq \frac{N\log\sqrt{2}}{\tau} + O(\tau)$. To achieve this upper
bound, the potential $U(\textbf{x})$ must ``flattened out'' to decrease $k_i$,
as described in Section \ref{sec:first-order}.

There are alternative models for coupled particles diffusing in a heat bath,
and there is no guarantee that even the qualitative conclusions here will hold
true when particle trajectories are modeled according to a second-order Langevin equation, for instance.

%%%%%%%%%%%%%%%%%%%%%%%%%%%%%%%%%%%%%%%%%%

\section{Conclusions}

Our calculations led to general formulae for the information anatomy of
stochastic equilibria in simple, familiar systems when the time discretization
was very small. We considered a first-order nonlinear Langevin equation with a
normalizable stationary distribution, invertible diffusion matrix, and
analytic drift. We do not expect the expressions in Section
\ref{sec:first-order} to hold for larger time discretizations, though Gaussian
approximations could be used to upper bound conditional entropies more
generally. We also considered first-order linear Langevin equations with
normalizable stationary distribution and noninvertible diffusion matrix in
Section \ref{sec:first-orderlinearLangevin}.

An important technical consideration is that the information anatomy of
Langevin stochastic dynamics is likely \textit{not} unique, just as the
pre-factors for the $(\epsilon,\tau)$-entropy rate of an Ornstein-Uhlenbeck
process depend on definition and approximation procedure \cite{Gasp93a,Gasp05a}.
However, based on results not shown here, we have reason to believe that
the qualitative scaling seen with drift and diffusion holds regardless of
approximation method. This parallels the way that $(\epsilon,\tau)$-entropy rate
estimates for an Ornstein-Uhlenbeck process all increase with diffusion
coefficient. That said, a complete understanding of how information anatomy
estimates vary with technique requires further study. We hope that our results
are sufficiently compelling to motivate further efforts.

With this caveat in mind, let's focus on qualitative rather than quantitative
conclusions. Even though the entropy rate is typically viewed as a measure of
randomness, some of that randomness is useful for prediction. That is the
bound information---shared between present and future but not contained in the
past---and we showed that it is sensitive only to drift. In contrast, we showed that the ephemeral information---information in the present useless
for predicting or retrodicting---is sensitive only to the diffusion. In short,
for stochastic equilibria the entropy rate consists of a quantity (ephemeral
information) that has to do with a process's inherent noisiness and a quantity
(bound information) that has only to do with the underlying process
regularities.

A key lesson is that information anatomy measures are sensitive to process
organization. Section \ref{sec:first-orderlinearLangevin} showed that the
information anatomy of linear Langevin dynamics changes discontinuously
whenever one of the diffusion coefficients vanishes.  This sensitivity to
underlying process structure could also be a feature rather than a defect. For instance, if we know that the underlying process is a first-order linear Langevin equation, then one could infer the dimension of the deterministically evolving state space by comparing known $\tau$-scaling relations in Section \ref{sec:first-order} with empirically determined scaling relations.

This brings us to discuss what was learned from the several example
applications.
Section \ref{sec:cusp_catastrophe} showed that the bound information picks up
different features than one finds in a dynamical phase diagram. In the
noiseless limit, $\bmu$ of the cusp catastrophe as a function of parameters
$r$ and $h$ is nondifferentiable on the line $h=0$ for $r\geq 0$, because the
location of the global minimum of the potential function changes
discontinuously across that contour.  Moreover, this is not related to the
bifurcation contour $h=\pm 2 r^{3/2} / 3\sqrt{3}$ \cite{Post78a} where the
number of equilibria changes from two to one or vice versa, which has no
apparent signature in the bound information. However, in these calculations,
we did not avoid the ``ultraviolet catastrophe''. We embraced it since we
could then evaluate the information anatomy for general nonlinear Langevin
equations by linearizing. If one evaluates the information anatomies of these
types of stochastic dynamics when the time discretization is not
infinitesimal, however, then signatures of dynamical phase transitions should
show up in the bound information as they do for the finite-time predictable
information or excess entropy \cite{Feld02b,Tche13a}.

Section \ref{sec:particles_heatbath} calculated the information anatomy of
coupled particles in a heat bath. Physicists are concerned primarily with
$H_0$, the entropy of a single measurement symbol, since its changes are
proportional to heat loss \cite{Kitt80a}. However, the point of this example
was that alternative information-theoretic quantities capture other behavioral
properties of particles diffusing in a heat bath. As an application of this
analysis it will be worth exploring how the information anatomy measures
reflect the trade-off between stable information storage and heat loss
in the context of Maxwell-like demons \cite{Land89a}.

To close our discussion of applications, we briefly mention the use of
information measures to express optimization principles that guide adaptive
agents. A Markov process's bound information has been used as an optimization
measure called the \emph{time-local predictive information} (TiPi)
\cite{Mart13a}. Moreover, the class of systems used there and for which TiPi
was calculated are exactly the first-order nonlinear Langevin dynamics analyzed
here. Due to the similarities in setup and approach, Appendix \ref{sec:TiPi}
compares alternative TiPi measures. Generally, an agent that wishes to maximize
its TiPi will be driven into unstable regions of the potential landscape on
which it diffuses. However, Appendix \ref{sec:TiPi} shows that the similarly
motivated, but alternative optimization measures lead to different adaptive
strategies. More investigation is required to compare such strategies to those
seen in biological agents before general principles of adaptive behavior can be
understood.

%%%%%%%%%%%%%%%%%%%%%%%%%%%%%%%%%%%%%%%%%%

\acknowledgements{Acknowledgments}

This material is based upon work supported by, or in part by, the U. S. Army Research Laboratory and the U. S. Army Research Office under contract number 
W911NF-13-1-0390. S.M. was funded by a National Science Foundation Graduate Student Research Fellowship and the U.C. Berkeley Chancellor's Fellowship.

%%%%%%%%%%%%%%%%%%%%%%%%%%%%%%%%%%%%%%%%%%

\conflictofinterests{Conflict of Interest}

The authors declare no conflict of interest.

%=================================================================
% References:
%=================================================================
% Use the following option to include external BibTeX files:
\bibliography{chaos}

\begin{thebibliography}{-------}
\providecommand{\natexlab}[1]{#1}

\bibitem[Walters(1982)]{Walt82a}
Walters, P.
\newblock {\em An Introduction to Ergodic Theory}; Vol.~79, {\em Graduate Texts
  in Mathematics}, Springer-Verlag: New York,
\newblock  1982.

\bibitem[Cover and Thomas(2006)]{Cove06a}
Cover, T.M.; Thomas, J.A.
\newblock {\em Elements of Information Theory}, second ed.; Wiley-Interscience:
  New York,
\newblock  2006.

\bibitem[Crutchfield and Young(1989)]{Crut88a}
Crutchfield, J.P.; Young, K.
\newblock Inferring Statistical Complexity.
\newblock {\em Phys. Rev. Let.} {\bf 1989},
\newblock {\em 63},~105--108.

\bibitem[Shalizi and Crutchfield(2001)]{Shal98a}
Shalizi, C.R.; Crutchfield, J.P.
\newblock Computational Mechanics: Pattern and Prediction, Structure and
  Simplicity.
\newblock {\em J. Stat. Phys.} {\bf 2001},
\newblock {\em 104},~817--879.

\bibitem[Crutchfield and Feldman(2003)]{Crut01a}
Crutchfield, J.P.; Feldman, D.P.
\newblock Regularities Unseen, Randomness Observed: Levels of Entropy
  Convergence.
\newblock {\em CHAOS} {\bf 2003},
\newblock {\em 13},~25--54.

\bibitem[James \em{et~al.}(2011)James, Ellison, and Crutchfield]{Jame11a}
James, R.G.; Ellison, C.J.; Crutchfield, J.P.
\newblock Anatomy of a Bit: {Information} in a Time Series Observation.
\newblock {\em CHAOS} {\bf 2011},
\newblock {\em 21},~037109.

\bibitem[Palmer \em{et~al.}(2013)Palmer, Marre, II, and Bialek]{Palm13a}
Palmer, S.E.; Marre, O.; II, M.J.B.; Bialek, W.
\newblock Predictive Information in a Sensory Population {\bf 2013}.
\newblock
\newblock arXiv:1307.0225.

\bibitem[Beer and Williams(2014)]{Beer14a}
Beer, R.D.; Williams, P.L.
\newblock Information Processing and Dynamics in Minimally Cognitive Agents.
\newblock {\em Cognitive Science} {\bf 2014},
\newblock p. in press.

\bibitem[Tononi \em{et~al.}(1998)Tononi, Edelman, and Sporns]{Tono98a}
Tononi, G.; Edelman, G.M.; Sporns, O.
\newblock Complexity and Coherency: {Integrating} Information in the Brain.
\newblock {\em Trends Cogn. Sci.} {\bf 1998},
\newblock {\em 2},~474--484.

\bibitem[Strelioff and Crutchfield(2013)]{Stre13a}
Strelioff, C.C.; Crutchfield, J.P.
\newblock Bayesian Structural Inference for Hidden Processes {\bf 2013}.
\newblock
\newblock Santa Fe Institute Working Paper 13-09-027, arXiv:1309.1392
  [stat.ML].

\bibitem[Sato \em{et~al.}(2005)Sato, Akiyama, and Crutchfield]{Sato04a}
Sato, Y.; Akiyama, E.; Crutchfield, J.P.
\newblock Stability and Diversity in Collective Adaptation.
\newblock {\em Physica D} {\bf 2005}, {\em 210},~21--57.
\newblock
\newblock Santa Fe Institute Working Paper 04-08-025;
  arXiv.org/abs/nlin/0408039.

\bibitem[Martius \em{et~al.}(2013)Martius, Der, and Ay]{Mart13a}
Martius, G.; Der, R.; Ay, N.
\newblock Information driven self-organization of complex robotics behaviors.
\newblock {\em PLoS One} {\bf 2013},
\newblock {\em 8},~e63400.

\bibitem[Varn \em{et~al.}(2002)Varn, Canright, and Crutchfield]{Varn02a}
Varn, D.P.; Canright, G.S.; Crutchfield, J.P.
\newblock Discovering Planar Disorder in Close-Packed Structures from {X-Ray}
  Diffraction: {Beyond} the Fault Model.
\newblock {\em Phys. Rev. B} {\bf 2002},
\newblock {\em 66},~174110--3.

\bibitem[Varn \em{et~al.}(2013)Varn, Canright, and Crutchfield]{Varn12a}
Varn, D.P.; Canright, G.S.; Crutchfield, J.P.
\newblock $\epsilon$-Machine spectral reconstruction theory: {A} direct method
  for inferring planar disorder and structure from {X}-ray diffraction studies.
\newblock {\em Acta. Cryst. Sec. A} {\bf 2013},
\newblock {\em 69},~197--206.

\bibitem[Crutchfield and Young(1990)]{Crut89e}
Crutchfield, J.P.; Young, K.
\newblock Computation at the Onset of Chaos.
\newblock  Entropy, Complexity, and the Physics of Information; Zurek, W., Ed.;
  Addison-Wesley: Reading, Massachusetts,  1990; Vol. VIII, {\em SFI Studies in
  the Sciences of Complexity},
\newblock pp. 223 -- 269.

\bibitem[Tchernookov and Nemenman(2013)]{Tche13a}
Tchernookov, M.; Nemenman, I.
\newblock Predictive Information in a Nonequilibrium Critical Model.
\newblock {\em J. Stat. Phys.} {\bf 2013},
\newblock {\em 153},~442--459.

\bibitem[Atmanspracher and Scheingraber(1991)]{Atma91a}
Atmanspracher, H.A.; Scheingraber, H.
\newblock {\em Information Dynamics}; Plenum: New York,  1991;
\newblock pp. 45 -- 60.

\bibitem[James \em{et~al.}(2014)James, Burke, and Crutchfield]{Jame13a}
James, R.G.; Burke, K.; Crutchfield, J.P.
\newblock Chaos Forgets and Remembers: {Measuring} Information Creation and
  Storage {\bf 2014}.
\newblock
\newblock Santa Fe Institute Working Paper 13-10-030, arXiv:1309.5504
  [nlin.CD].

\bibitem[Lizier \em{et~al.}(2010)Lizier, Prokopenko, and Zomaya]{Lizi10a}
Lizier, J.; Prokopenko, M.; Zomaya, A.
\newblock Information modification and particle collisions in distributed
  computation.
\newblock {\em CHAOS} {\bf 2010},
\newblock {\em 20},~037109.

\bibitem[Flecker \em{et~al.}(2011)Flecker, Alford, Beggs, Williams, and
  Beer]{Flec11a}
Flecker, B.; Alford, W.; Beggs, J.M.; Williams, P.L.; Beer, R.D.
\newblock Partial Information Decomposition as a Spatiotemporal Filter.
\newblock {\em CHAOS} {\bf 2011},
\newblock {\em 21},~037104.

\bibitem[Moss and McClintock()]{Moss89a}
Moss, F.; McClintock, P.V.E.
\newblock {\em Noise in Nonlinear Dynamical Systems}; Vol.~1,
\newblock Cambridge University Press.

\bibitem[Shraiman \em{et~al.}(1981)Shraiman, Wayne, and Martin]{Shra81}
Shraiman, B.; Wayne, C.E.; Martin, P.C.
\newblock Scaling Theory for Noisy Period-Doubling Transitions to Chaos.
\newblock {\em Phys. Rev. Lett.} {\bf 1981},
\newblock {\em 46},~935.

\bibitem[Crutchfield \em{et~al.}(1981)Crutchfield, Nauenberg, and
  Rudnick]{Crut81}
Crutchfield, J.P.; Nauenberg, M.; Rudnick, J.
\newblock Scaling for External Noise at the Onset of Chaos.
\newblock {\em Phys. Rev. Lett.} {\bf 1981},
\newblock {\em 46},~933.

\bibitem[Yeung(2008)]{Yeun08a}
Yeung, R.W.
\newblock {\em Information Theory and Network Coding}; Springer: New York,
\newblock  2008.

\bibitem[Gaspard and Wang(1993)]{Gasp93a}
Gaspard, P.; Wang, X.J.
\newblock Noise, Chaos, and ($\epsilon,\tau$)-Entropy Per Unit Time.
\newblock {\em Physics Reports} {\bf 1993},
\newblock {\em 235},~291--343.

\bibitem[Gaspard(2005)]{Gasp05a}
Gaspard, P.
\newblock Brownian Motion, Dynamical Randomness, and Irreversibility.
\newblock {\em New Journal of Physics} {\bf 2005},
\newblock {\em 7},~77--90.

\bibitem[Lecomte \em{et~al.}(2007)Lecomte, Appert-Rolland, and van
  Wijland]{Leco07a}
Lecomte, V.; Appert-Rolland, C.; van Wijland, F.
\newblock Thermodynamic Formalism for Systems with Markov Dynamics.
\newblock {\em J. Stat. Phys.} {\bf 2007},
\newblock {\em 127},~51--106.

\bibitem[Ellison \em{et~al.}(2009)Ellison, Mahoney, and Crutchfield]{Crut08b}
Ellison, C.J.; Mahoney, J.R.; Crutchfield, J.P.
\newblock Prediction, Retrodiction, and the Amount of Information Stored in the
  Present.
\newblock {\em J. Stat. Phys.} {\bf 2009},
\newblock {\em 136},~1005--1034.

\bibitem[Crutchfield \em{et~al.}(2009)Crutchfield, Ellison, and
  Mahoney]{Crut08a}
Crutchfield, J.P.; Ellison, C.J.; Mahoney, J.R.
\newblock Time's Barbed Arrow: {Irreversibility}, Crypticity, and Stored
  Information.
\newblock {\em Phys. Rev. Lett.} {\bf 2009}, {\em 103},~094101.
\newblock
\newblock SFI Working Paper 09-02-002; arxiv.org:0902.1209
  [cond-mat.stat-mech]; DOI 10.1103/PhysRevLett.103.094101.

\bibitem[Crutchfield and Feldman(1997)]{Crut97a}
Crutchfield, J.P.; Feldman, D.P.
\newblock Statistical Complexity of Simple One-Dimensional Spin Systems.
\newblock {\em Phys. Rev. E} {\bf 1997},
\newblock {\em 55},~R1239--R1243.

\bibitem[Poston and Stewart(1978)]{Post78a}
Poston, T.; Stewart, I.
\newblock {\em Catastrophe Theory and Its Applications}; Pitman: London,
\newblock  1978.

\bibitem[Feldman and Crutchfield(2003)]{Feld02b}
Feldman, D.P.; Crutchfield, J.P.
\newblock Structural Information in Two-Dimensional Patterns: Entropy
  Convergence and Excess Entropy.
\newblock {\em Phys. Rev. E} {\bf 2003},
\newblock {\em 67},~051103.

\bibitem[Kittel and Kroemer(1980)]{Kitt80a}
Kittel, C.; Kroemer, H.
\newblock {\em Thermal Physics}, second ed.; W. H. Freeman: New York,
\newblock  1980.

\bibitem[Landauer(1989)]{Land89a}
Landauer, R.
\newblock Dissipation and Noise Immunity in Computation, Measurement, and
  Communication.
\newblock {\em J. Stat. Phys.} {\bf 1989},
\newblock {\em 54},~1509--1517.

\bibitem[Lohr(2009)]{Lohr09a}
Lohr, W.
\newblock Properties of the Statistical Complexity Functional and Partially
  Deterministic HMMs.
\newblock {\em Entropy} {\bf 2009},
\newblock {\em 11},~385--401.

\bibitem[Risken(2007)]{Risk07a}
Risken, H.
\newblock In {\em The Fokker-Planck Equation: Methods of Solution and
  Applications};  2007;
\newblock pp. 32--58.

\bibitem[Drozdov and Morillo(1996)]{Droz96a}
Drozdov, A.N.; Morillo, M.
\newblock Expansion for the Moments of a Nonlinear Stochastic Model.
\newblock {\em Phys. Rev. Lett.} {\bf 1996},
\newblock {\em 77},~3280.

\bibitem[Crutchfield \em{et~al.}(2010)Crutchfield, Ellison, Mahoney, and
  James]{Crut10a}
Crutchfield, J.P.; Ellison, C.J.; Mahoney, J.R.; James, R.G.
\newblock Synchronization and Control in Intrinsic and Designed Computation:
  {An} Information-Theoretic Analysis of Competing Models of Stochastic
  Computation.
\newblock {\em CHAOS} {\bf 2010},
\newblock {\em 20},~037105.

\end{thebibliography}
\bibliographystyle{mdpi}

\appendix

\section{Information Anatomy of a Markov Process}
\label{sec:Markov}

If the system at hand is Markovian, then the information anatomy simplifies
tremendously since one need only consider single time steps into the future and
into the past. As a result, many of the Markovian formulae are special cases of
those developed in \cite{Jame11a} for more complex processes, but are derived here for completeness.

For notational ease, we use the discrete-time notation in which $X_{t:t'}$ is
the random variable of measurements $X_t,X_{t+1},...,X_{t'-1}$. For a
Markovian process the immediately preceding observation ``shields'' the future from the past:
\begin{align*}
\Prob(X_{n} = x_{n}|X_{-m:n} = x_{-m:n}) = \Prob(X_{n}=x_{n}|X_{n-1}=x_{n-1})
  ~.
\end{align*}
And, it becomes relatively easy to calculate the information
anatomy measures, since the sequence probabilities simplify:
\begin{align*}
\Prob(X_{-m:n+1} = x_{-m:n+1})
  = \Prob(X_{-m}=x_{-m}) \prod_{k=-m}^{n-1} \Prob(X_{k+1}=x_{k+1}|X_{k}=x_{k})
  ~.
\end{align*}
For example, the entropy rate becomes:
\begin{align*}
  \hmu & = H[X_0|X_{:0}] \\
       & = H[X_0|X_{-1}]
  ~.
\end{align*}
Moreover, all information shared between the past and future goes through the present:
\begin{align*}
\sigma_{\mu} & = I[X_{:0};X_{1:}|X_0] \\
  & = H[X_{1:}|X_0] - H[X_{1:}|X_{:1}] \\
  & = H[X_{1:}|X_0] - H[X_{1:}|X_0] \\
  & = 0
  ~.
\end{align*}
Finally, the mutual information between the present and the future conditioned
on the past (bound information) is:
\begin{align*}
\bmu & = I[X_0;X_{1:}|X_{:0}] \\
  & = H[X_{1:}|X_{:0}] - H[X_{1:}|X_{:1}] \\
  & = H[X_{1}|X_{:0}] - H[X_{1}|X_{:1}]
	+ H[X_{2:}|X_{:0},X_{1}] - H[X_{2:}|X_{:2}] \\
  & = H[X_{1}|X_{-1}] - H[X_{1}|X_{0}] \\
  & = I[X_1;X_0|X_{-1}]
  ~.
\end{align*}
This equality is evident from the information diagram of Figure
\ref{fig1}\subref{fig1B}.
The other information anatomy measures follow from $\bmu$ and $\hmu$ via
identities given in Section \ref{sec:Background}:
\begin{align*}
\rmu & = h_{\mu} - b_{\mu} \\
     & = H[X_0|X_{-1}] - I[X_1;X_0|X_{-1}] ~\text{and} \\
\qmu & = H[X_0] - h_{\mu} - b_{\mu} \\
     & = H[X_0] - H[X_0|X_{-1}] - I[X_1;X_0|X_{-1}]
  ~.
\end{align*}
The excess entropy follows as the sum:
\begin{align*}
\EE & = \sigma_{\mu} + q_{\mu} + b_{\mu} \\
    & = H[X_0] - h_{\mu} \\
    & = H[X_0] - H[X_0|X_{-1}]
  ~.
\end{align*}
As stated in Section \ref{sec:Background}, to normalize these measures as
rates (entropies per unit time rather than per measurement), we simply divide the above above by the time discretization $\tau$:
\begin{align*}
\hmu (\tau) & = \frac{H[X_0|X_{-\tau}]}{\tau} \\
\bmu (\tau) & = \frac{I[X_{\tau};X_0|X_{-\tau}]}{\tau} \\
\rmu (\tau) & = \frac{H[X_0|X_{-\tau}] - I[X_{\tau};X_0|X_{-\tau}]}{\tau} \\
\qmu (\tau) & = \frac{H[X_0] - H[X_0|X_{-\tau}] - I[X_{\tau};X_0|X_{-\tau}]}{\tau}
  ~.
\end{align*}

If the system is Markovian, one only needs the joint distribution of three
successive measurements to calculate the anatomy of a bit. Thus, the formulae
derived here also can be used as time-local measures for nonstationary dynamics
despite the subtleties of defining a measure over bi-infinite time series in
general \cite{Lohr09a}.

\section{Statistical Complexity is the Entropy of a Measurement}
\label{sec:causal_states}

The statistical complexity $\Cmu$ is the entropy of the probability distribution over
causal states. Causal states themselves are groupings of pasts that are
partitioned according to the predictive equivalence relation $\sim_{\epsilon}$
\cite{Shal98a}:
\begin{align*}
\past \sim_{\epsilon} \past' \Leftrightarrow
  \Prob (\Future | \Past = \past) = \Prob (\Future | \Past = \past')
  ~.
\end{align*}
Although causal states are difficult to determine for general complex
processes, they are particularly easy for Markov processes. Recall that a
Markov process is defined by single-time step shielding:
\begin{align*}
\Prob (\MS_{0:}| \MS_{:0})
  = \Prob(\MS_{0} | \MS_{-\tau}) \Prob(\MS_{1:} | \MS_0)
  ~.
\end{align*}
It follows that:
\begin{align*}
\Prob(\Future|\Past = \past) = \Prob(\Future|\Past = \past')
  \Leftrightarrow
  \Prob(\MS_0|\MS_{-1} = \ms_{-1}) = \Prob(\MS_0 | \MS_{-1} = \ms'_{-1})
  ~.
\end{align*}
Therefore, for a Markov process, groupings of pasts in which only the last
measurement is recorded constitutes at least a prescient partition. Since:
\begin{align*}
\Prob(\MS_0|\MS_{-1} = \ms) = \Prob(\MS_0 | \MS_{-1} = \ms')
  \Leftrightarrow \ms = \ms'
  ~,
\end{align*}
we conclude that the causal states are simply groupings of pasts with the
same last measurement: $\epsilon(\ms_{:0}) = \ms_{-1}$. The causal state space
$\mathcal{S}$ is isomorphic to the alphabet of the process $\mathcal{A}$ and
the statistical complexity is the entropy of a single measurement:
$\Cmu = H[\MS_0]$.

First-order Langevin equations generate Markovian time series.  Our claim,
then, is that the stochastic differential equations considered here produce
time series for which:
\begin{align*}
\Prob(\MS_0|\MS_{-\tau} = \ms) = \Prob(\MS_0 | \MS_{-\tau} = \ms')
  \Leftrightarrow \ms = \ms'
  ~.
\end{align*}
So, the causal states are isomorphic to the present measurement $\MS_0$ and
the statistical complexity is $\Cmu = H[\MS_0]$. Implicit in these calculations
is an assumption that the transition probabilities $\Prob(\MS_0|\MS_{-\tau})$
for a given stochastic differential equation exist and are unique.

For intuition, consider a linear Langevin dynamics for an Ornstein-Uhlenbeck process:
\begin{align*}
dX_t = A dt+ BX_t dt + \sqrt{D} dW_t
  ~.
\end{align*}
As described in Appendix \ref{sec:linearLangevin} and many other places, e.g., \cite{Moss89a}, the transition probability density
$\Prob(\MS_t|\MS_0=\ms)$ is a Gaussian:
\begin{align*}
\Prob (\MS_t|\MS_0=\ms)
  \sim \mathcal{N}
  \left( e^{B t} \ms + e^{Bt} \int_0^t e^{-B t'} A dt',
  \int_0^t e^{B t'} D e^{B^{\top}t'} dt' \right)
  ~.
\end{align*}
For $\Prob(\MS_t|\MS_0=\ms) = \Prob(\MS_t|\MS_0=\ms')$ the means and variances
of the above probability distribution must match. Meaning that $e^{Bt} \ms =
e^{Bt} \ms' \Rightarrow \ms = \ms'$.  Therefore, for an Ornstein-Uhlenbeck
process, the causal states are indeed isomorphic to the present measurement
and the statistical complexity is $H[\MS_0]$. The key here is that although $\Prob(\MS_t|\MS_0=\ms)$ may quickly forget its initial condition $\ms$, for any finite-time discretization, the transition probability $\Prob(\MS_t|\MS_0=\ms)$ still depends on $\ms$.

In the more general case, we have a nonlinear Langevin equation:
\begin{align*}
d \MS_t = -D\nabla U dt + \sqrt{2D} dW_t
  ~,
\end{align*}
where the stationary distribution $\rho_{eq}$ exists and is normalizable.
Our goal is to show that if
$\Prob(\MS_t|\MS_0 = \ms)=\Prob(\MS_t|\MS_0=\ms')$, then $\ms=\ms'$.
The transition probability $\Prob(\MS_t=\ms|\MS_0=\ms')$ is a solution to
the corresponding Fokker-Planck equation:
\begin{align*}
\frac{\partial \rho(x,t)}{\partial t} = -\nabla \cdot (\mu(x) \rho(x,t)) + D \nabla^2 \rho(x,t)
  ~,
\end{align*}
with initial condition $\rho(x,0)=\delta(x-x')$.  
%If the solution to the corresponding Fokker-Planck equation is unique given the initial conditions, then $P(X_t|X_0 = x) = P(X_t|X_0 = x')$ must imply that 
%The transition probability $P(X_t|X_0=x)$ is well-defined \cite{Risk07a}.
%Although existence and uniqueness to Fokker-Planck equations usually requires
%that the drift be suitably bounded over the entire state space by a function
%linear in $||x||$, e.g. as described in Section $3$ of \cite{Bris08a}, if the stationary distribution is normalizable, then the drift will be bounded over the region of the state space in which the position probability is not infinitesimally small.
As in \cite{Risk07a}, we can use an eigenfunction expansion to show that
$\rho(x,t|x',0)$ cannot equal $\rho(x,t|x'',0)$ unless $x'=x''$ for finite
time $t$. Therefore,
$\Prob(\MS_t|\MS_0=x') = \Prob(\MS_t|\MS_0 =x'') \Rightarrow x'=x''$.
This implies that the causal states are again isomorphic to the present
measurement and the statistical complexity is $\Cmu = H[\MS_0]$.

To summarize, this application of computational mechanics
\cite{Crut88a,Shal98a} to Langevin stochastic dynamics shows that the entropy
of a single measurement is also the process's statistical complexity $\Cmu$.
Recall that the latter is the entropy of the probability distribution over the
causal states, which in turn are groupings of pasts that lead to equivalent
predictions of future behavior. So, for the stochastic differential equations
considered here, their causal states simply track the last measured position.
What the information anatomy analysis reveals, then, is that not all of the
information required for optimal prediction is predictable information about
the future. In other words, Langevin stochastic dynamics are inherently cryptic
\cite{Crut08a,Crut08b}. Unfortunately, as is so often the case, the necessary
and the apparent come packaged together and cannot be teased apart without
effort.

\section{Approximating the Short-time Propagator Entropy}
\label{sec:Greens}

The study of stochastic differential equations and short-time propagator
approximations is mathematically rich and, as noted in the introduction,
the application to nonlinear diffusion has a long history \cite{Moss89a}.
What follows is a brief sketch, not a rigorous proof, that likely glosses
over important pathological cases.

Consider the nonlinear Langevin equation:
\begin{align}
\frac{dx}{dt} = -D\nabla U(x) + \eta(t)
\label{eq:App2_setup1}
  ~.
\end{align}
with driving noise satisfying $\langle \eta(t)\rangle = 0$ and
$\langle \eta(t) \eta^{\top}(t') \rangle = D \delta(t-t')$, where
$\det D \neq 0$. Let $p(x|x')$ be the transition probability
$\Prob(\MS_t=x|\MS_0 = x')$ for the system in Eqn.~\ref{eq:App2_setup1}. 
From arguments in \cite{Risk07a}, it exists and is uniquely defined when
the stationary distribution is normalizable. Let $q(x|x')$ be a Gaussian
with the same mean and variance as $p(x'|x)$.

We show that $H[p] = H[q] + o(\tau)$ where
$H[p]=-\int p(x|x') \log p(x|x') dx$ and $H[q]=-\int q(x|x')\log q(x|x') dx$.
Note that here and in the following we suppress notation for the dependence of
these quantities on $x'$, using the shorthand $H[p] \equiv H[p|X' = x']$ and
the like. First, consider:
\begin{align*}
D_{KL}[p||q] &= \int p(x|x') \log \frac{p(x|x')}{q(x|x')} dx \\
  &= \int p(x|x') \log p(x|x') dx - \int p(x|x') \log q(x|x') dx \\
  &= -H[p] - \int p(x|x') \log q(x|x') dx
  ~.
\end{align*}
Since $q(x|x')$ is the Maximum Entropy distribution consistent with the mean
and the variance of $p(x|x')$, averages of $\log q(x|x')$ with respect to $p$
are the same as those with respect to $q$. Specifically, if $\bar{x}$ is the
mean:
\begin{align*}
\bar{x} =\int x p(x|x') dx = \int x q(x|x') dx
\end{align*}
and if $C(x')$ is the variance:
\begin{align*}
C(x') & = \int (x-\bar{x}) (x-\bar{x})^{\top} q(x|x') dx \\
      & = \int (x-\bar{x}) (x-\bar{x})^{\top} p(x|x') dx
  ~,
\end{align*}
then $q$ is the normal distribution consistent with that mean and variance:
\begin{align*}
q(x|x') = \frac{1}{\sqrt{2\pi |\det C(x')|}}
  \exp \left(-\frac{1}{2} (x-\bar{x})^{\top} C(x')^{-1} (x-\bar{x})\right)
  ~.
\end{align*}
From this, we derive:
\begin{align*}
\int p(x|x') \log q(x|x') dx &=
  \int p(x|x')
  \log \frac{e^{-\frac{1}{2}(x-\bar{x})^{\top}C(x')^{-1}(x-\bar{x})}}
  {\sqrt{2\pi |\det C(x')|}} \\
  & = -\frac{1}{2}\int (x-\bar{x})^{\top} C(x')^{-1}(x-\bar{x})) p(x|x') dx
  -\log\sqrt{2\pi |\det C(x')|}
\end{align*}
Since the mean and variance for $p$ and $q$ are consistent, we have:
\begin{align*}
\int p(x|x') \log q(x|x') dx & =
  -\frac{1}{2}\int (x-\bar{x})^{\top} C(x')^{-1}(x-\bar{x})) q(x|x') dx
  \nonumber \\
  & -\log\sqrt{2\pi |\det C(x')|}  \\
  &= \int q(x|x') \log q(x|x') dx \\
  &= -H[q]
\end{align*}
and, thus:
\begin{align*}
D_{KL}[p||q] &= H[q] - H[p]
  ~.
\end{align*}

Hence, if we can show that $D_{KL}[p||q]$ is $o(\tau)$, we will have shown
that $H[p] = H[q]+o(\tau)$. Finally, we also want to show that $H[q]$ can be
determined to $o(\tau)$ from the linearized Langevin equation:
\begin{align*}
\frac{dx}{dt} = \mu(x')
  + \frac{\partial \mu(x)}{\partial x}|_{x=x'} (x-x') + \eta(t)
  ~.
%\label{eq:App2_setup2}
\end{align*}
A moment expansion will show that the moments of $q$ can be determined to $o(\tau)$ from this linearized Langevin equation.

Our strategy is to construct a series expansion for the moments of $p$ in the
timescale $\tau$, as in \cite{Droz96a}. Immediately, with that statement, we
run into a problem. Moments do not uniquely specify a distribution unless an
additional condition (e.g., Carleman's condition) is satisfied. 
%Most of the proofs for existence and uniqueness of solutions to Fokker-Planck
%equations require that the drift be bounded by $||\mu(x)|| \leq M(1+||x||)$
%for some $M<\infty$ \cite{Bris08a}, but this we have only specified that the drift be analytic.  And we have concentrated on situations in which this boundedness is explicitly violated, e.g. the cusp catastrophe in Section \ref{sec:cusp_catastrophe}!
However, we are interested in approximating the entropy of the transition probability, rather than approximating the transition probability itself.
The Kullback-Liebler divergence is invariant to changes in the coordinate
system and, for reasons that become apparent later, it is useful to move to the
parametrization $z= (x-\bar{x})/\sqrt{t}$. In a slight abuse of notation, $p(z|x')$ and $q(z|x')$ will be used to denote the re-parametrized distributions $p(x|x')$ and $q(x|x')$.
If we could show that all moments of $p(z|x')$ and $q(z|x')$ differ by a quantity that is at most of $O(\tau^{3/2})$, it would follow that $p(z|x') = q(z|x') + \tau^{3/2} \delta q$ where $\delta q$ is at most of $O(1)$ in $\tau$. From that it would follow that $D_{KL}[q+\tau^{3/2}\delta q||q] = (\tau^{3/2})^2 \mathcal{I}[q]$ where $\mathcal{I}[q]$ is the Fisher information of a Gaussian (and hence bounded) and that $H[p]=H[q]$ to $o(\tau)$.

For intuition and simplicity, we start with the one-dimensional example.
This is similar in flavor to the approach in \cite{Droz96a}, but our point
differs---we wish to understand how well we can approximate the full system
with a linearized drift term. The stochastic differential equation for
$x\in\mathbb{R}$ is:
\begin{align*}
\frac{dx}{dt} & = \mu(x) + \eta(t) ~,
\end{align*}
with noise as above. The mean $\langle x\rangle$ evolves according to:
\begin{align*}
\frac{d\langle x\rangle}{dt} = \langle \mu(x)\rangle
  ~.
\end{align*}
Using an Ito discretization scheme:
\begin{align*}
x(t+\Delta t) = x(t) + \mu(x(t))\Delta t + d\eta(t)
  ~,
\end{align*}
where $d\eta(t)\sim\mathcal{N}(0,D\Delta t)$, we have:
\begin{align}
x(t+\Delta t) - \langle x(t+\Delta t)\rangle
  & = x(t) - \langle x(t)\rangle +
  (\mu(x(t))-\langle \mu(x(t))\rangle )\Delta t  + d\eta(t)
  ~.
\label{eq:00}
\end{align}
From these, we derive evolution equations for the moments
$\langle (x-\langle x\rangle)^n\rangle$ for $n\geq 2$:
\begin{align}
\frac{d\langle (x-\langle x\rangle)^n\rangle}{dt}
  & = \lim_{\Delta t\rightarrow 0}
  \Big\langle \frac{(x(t+\Delta t)-\langle x(t+\Delta t)\rangle)^n}{\Delta t}
  - \frac{(x(t)-\langle x(t)\rangle)^n}{\Delta t} \Big\rangle
  ~.
\label{eq:001}
\end{align}
Substituting Eqn.~\ref{eq:00} into the above and simplifying leads to:
\begin{align}
\frac{d\langle (x-\langle x\rangle)^n\rangle}{dt}
  & = n \langle (x-\langle x\rangle)^{n-1}
  (\mu(x)-\langle \mu(x)\rangle)\rangle
  + {n\choose 2} D \langle (x-\langle x\rangle)^{n-2}\rangle
  ~.
  \label{eq:002}
\end{align}
Now, we re-express:
\begin{align*}
\mu(x) = \mu(x') + \mu'(x') (x-x') + \delta(x,x') (x-x')^2
  ~,
\end{align*}
where $\delta$ is at most $O(1)$ in $x-x'$. Then:
\begin{align*}
\frac{d\langle (x-\langle x\rangle)^n\rangle}{dt}
  & = n \mu'(x') \langle (x-\langle x\rangle)^{n-1} (x-\langle x\rangle) \rangle
  + n\langle (x-\langle x\rangle)^{n-1} (\delta-\langle \delta \rangle)\rangle \\
  & \quad\quad\quad + {n\choose 2} D \langle (x-\langle x\rangle)^{n-2}\rangle\\
  & = n \mu'(x') \langle (x-\langle x\rangle)^n\rangle
  + n\langle (x-\langle x\rangle)^{n+1} (\delta-\langle \delta \rangle)\rangle
  + {n\choose 2} D \langle (x-\langle x\rangle)^{n-2}\rangle
  ~.
\end{align*}
When $\mu'(x')=0$ and $\delta = 0$ the Green's function is a Gaussian with zero
mean and variance $Dt$, so that
$\langle (x-\langle x\rangle)^n\rangle \propto (Dt)^{n/2}$. Inspired by this
base case, we consider the moments of the variable
$z = (x-\langle x\rangle)/\sqrt{Dt}$:
\begin{align}
\frac{d\langle z^n\rangle}{dt} & = -\frac{n}{2t}
  \langle z^n\rangle + (Dt)^{-n/2} \frac{d (x-\langle x\rangle)^n}{dt}
  \nonumber \\
  & = -\frac{n}{2t} \langle z^n\rangle + n\mu'(x') \langle z^n\rangle
  + n \sqrt{Dt} \langle z^{n+1} (\delta-\langle\delta\rangle) \rangle
  + {n \choose 2} \frac{\langle z^{n-2} \rangle}{t}
  ~.
\label{eq:0}
\end{align}
We expand $\langle z^n\rangle$ in terms of $t$, since we are interested in the small-$t$ limit:
\begin{equation}
\langle z^n\rangle = C_n + \alpha_n\sqrt{t} + \beta_n t + \gamma_n t^{3/2} + O(t^2)
  ~.
\label{eq:1}
\end{equation}
In terms of these coefficients, we have:
\begin{equation}
\frac{d\langle z^n\rangle}{dt} = \frac{\alpha_n}{2\sqrt{t}} + \beta_n + \frac{3}{2} \gamma_n \sqrt{t} + O(t)
  ~.
\label{eq:2}
\end{equation}
Substituting Eqns.~\ref{eq:1}-\ref{eq:2} into Eqn.~\ref{eq:0} and matching
$O(1/t)$ terms, $O(1/\sqrt{t})$ terms, and so on, yields:
\begin{align}
0 & = -\frac{n}{2} C_n + {n\choose 2} C_{n-2} ~, \label{eq:11}\\
  \frac{\alpha_n}{2} & = -\frac{n}{2}\alpha_n + {n\choose 2} \alpha_{n-2}
  ~, \text{and} \\
  \beta_n & = -\frac{n}{2} \beta_n + n\mu'(x') C_n + {n\choose 2} \beta_{n-2}
  ~,
\label{eq:22}
\end{align}
for $O(1/t)$, $O(1/\sqrt{t})$, and $O(1)$, respectively. Note that none of $C_n$, $\alpha_n$, or $\beta_n$ have information about $\delta$, which encapsulates higher-order nonlinearities of the drift. The $O(\sqrt{t})$ term finally has information about $\delta$:
\begin{align*}
\frac{3}{2}\gamma_n = -\frac{n}{2}\gamma_n + n\mu'(x') \alpha_n + n\sqrt{Dt} \delta (x=x') C_n + {n\choose 2} \gamma_{n-2}.
\end{align*}
Interestingly, this implies that any dependencies of the moments on $\delta$
are $O(t^{3/2})$, at most. Eqns.~\ref{eq:11}-\ref{eq:22} can be solved with the following initial conditions:
\begin{align*}
\langle z^0\rangle = 1 \rightarrow C_0 = 1,~\alpha_0 = 0,~\beta_0 = 0
\end{align*}
and, by construction:
\begin{align*}
\langle z^1 \rangle = 0 \rightarrow C_1 = 0 ,~\alpha_1 = 0,~\beta_1 = 0
  ~.
\end{align*}
Then, $C_n = \alpha_n = \beta_n = 0$ for $n$ odd, and $\alpha_n = 0$ for $n$
even as well. Some algebra shows that:
\begin{align*}
C_n & = \begin{cases}
  \frac{n!}{(n/2)! 2^{n/2}} & n~\text{even} \\ 0 & n~\text{odd}
  \end{cases} \\
  \alpha_n & = 0 \\
  \beta_n & = \begin{cases}
  \frac{n}{2} C_n \mu'(x') & n~\text{even} \\ 0 & $n$~\text{odd}
  \end{cases}
  ~.
\end{align*}
A Gaussian with mean $0$ and variance $C_2+\alpha_2\sqrt{t}+\beta_2 t =
1+\mu'(x') t$ would also have $C_n = \alpha_n = \beta_n = 0$ for $n$ odd,
$\alpha_n = 0$ for $n$ even, and $\langle z^n\rangle_{q} = C_n (1+\beta_2
t)^{n/2} = C_n + \frac{n}{2} C_n \mu'(x') t + O(t^2)$.  Thus, the moments
$z^n$ of $p(z|x')$ are consistent with the moments of $q(z|x')$ to
$O(t^{3/2})$. And, those moments are consistent with the moments of the linearized Langevin equation to $O(t^2)$. From prior logic, $H[p]$ can be approximated to $O(t^2)$ by $\frac{1}{2}\log (2\pi e |Dt + \mu'(x') D t^2|)$.

The $n$-dimensional case follows the same principle, but the calculations
are more arduous. We start with the stochastic differential equation for
$x \in \mathbb{R}^n$:
\begin{align*}
\frac{dx}{dt} = \mu(x) + \eta(t) ~,
\end{align*}
with the noise as before. The initial condition is $x(t=0)=x'$. Since we are
interested not only in whether the distribution is effectively Gaussian, but
also in how important the nonlinearities of $\mu(x)$ are, we re-express
$\mu(x)$ as:
\begin{align*}
\mu(x) = \mu(x')+A(x') (x-x') + f(x) ~,
\end{align*}
where $A_{ij}(x') = \partial \mu_j / \partial x_i$:
\begin{align}
f_i(x) = \sum_{j,k} \delta_{ijk} (x_j-x'_j)(x_k-x'_k)
  ~,
\label{eq:f00}
\end{align}
and $\delta_{ijk}$ is at most of $O(1)$ in $||x-x'||$.
The evolution equation for the means is:
\begin{align*}
\frac{d\langle x\rangle}{dt}
  = \mu(x') + A(x') (\langle x\rangle-x') + \langle f(x) \rangle
  ~.
\end{align*}
Using an Ito discretization scheme with time step $\Delta t$:
\begin{align*}
x(t+\Delta t) = x(t)
  + \mu(x') \Delta t + A(x') (x-x') \Delta t + f(x) \Delta t + d\eta(t)
  ~,
\end{align*}
where $d\eta(t) \sim \mathcal{N}(0,D\Delta t)$. From this, we find evolution
equations for the moments of $x$. As before, we subtract the mean:
\begin{align}
x(t+\Delta t)-\langle x(t+\Delta t)\rangle
  = x(t) & - \langle x(t)\rangle + A(x') (x(t)-\langle x(t)\rangle) \Delta t
  \nonumber \\
  & + (f(x)-\langle f(x)\rangle) \Delta t + d\eta(t)
  ~.
\label{eq:000}
\end{align}
For notational ease, let $\sigma(1),...,\sigma(m)$ be a list of integers in the set $\{1,...,n\}$ where $n$ is the dimension of $x$; repeats are allowed.  We want an evolution equation for $\text{Cov}(x_{\sigma(1)},...,x_{\sigma(m)})$:
\begin{align*}
\frac{d}{dt} \text{Cov}(x_{\sigma(1)},...,x_{\sigma(m)})
  & = \frac{d}{dt}
  \left\langle \prod_{i=1}^m (x_{\sigma(i)}-\langle x_{\sigma(i)}\rangle)
  \right\rangle
  ~.
\end{align*}
Using Eqn.~\ref{eq:000} and steps similar to those outlined in
Eqns.~\ref{eq:001} and \ref{eq:002}, we find that:
\begin{align}
\frac{d}{dt} \text{Cov}(x_{\sigma(1)},...,x_{\sigma(m)})
  & = \sum_{i=1}^m \sum_{k=1}^n A_{ik}
  \text{Cov}(x_{\sigma(k)},x_{\sigma(j),~j\neq i}) \nonumber \\
  & + \sum_{i=1}^m \Big\langle (f_{\sigma(i)}(x)-\langle f_{\sigma(i)}(x)\rangle) \prod_{j\neq i}
  (x_{\sigma(j)}-\langle x_{\sigma(j)}\rangle)\Big\rangle \nonumber \\
  & + \sum_{i,j=1}^m \text{Cov}(x_{\sigma(k):k\neq i,j})
  ~.
\end{align}
%Using Eqn.~\ref{eq:f00},
%\begin{eqnarray}
%\frac{d\text{Cov}(x_{\sigma(1)},...,x_{\sigma(n)})}{dt} &=& \sum_{i=1}^m \sum_{k=1}^n A_{ik} \text{Cov}(x_{\sigma(k)},x_{\sigma(j),~j\neq i}) \nonumber \\
%&& + \sum_{i,j,k} \langle (\delta_{ijk}-\langle \delta_{ijk}\rangle) (x_{\sigma(j)}-\langle x_{\sigma(j)}\rangle)(x_{\sigma(k)}-\langle x_{\sigma(k)}\rangle) \prod_{l\neq i,j,k} (x_{\sigma(l)}-\langle x_{\sigma(l)}\rangle)\rangle \nonumber \\
%&& + \sum_{i,j=1}^m \text{Cov}(x_{\sigma(k):k\neq i,j})
%\end{eqnarray}
The notation $\text{Cov}(x_{\sigma(k):k\neq i,j})$ means the covariance of the variables $x_{\sigma(k)}$ for all $k$ in the integer list $1,...,m$ with the restriction that we ignore $k=i$ and $k=j$.
We have a base case: When $f=0$, $A=0$, and $D_{ij} = D \delta_{i,j}$, the Green's function is a Gaussian with variance $\propto \sqrt{t}$. So again, we switch to variable $z= \frac{x-\langle x\rangle}{\sqrt{t}}$ and calculate its covariance evolution, similarly to Eqn.~\ref{eq:2}, where we employ Eqn.~\ref{eq:f00} to find the appropriate $t$ scaling of the nonlinear $f$ term:
\begin{align}
\frac{d\text{Cov}(z_{\sigma(1)},...,z_{\sigma(m)})}{dt}
  & = - \frac{m}{2t} \text{Cov}(z_{\sigma(1)},...,z_{\sigma(m)})
  + \sum_{i=1}^m \sum_{k=1}^n A_{ik}
  \text{Cov}(z_{\sigma(k)},z_{\sigma(j),~j\neq i}) \nonumber \\
  & + \sqrt{t} \sum_{i,j,k} \langle \delta_{ijk}
  z_{\sigma(j)} z_{\sigma(k)} \prod_{l\neq i} z_{\sigma(l)} \rangle
  + \frac{1}{t} \sum_{i,j} D_{ij} \text{Cov}(z_{\sigma(k):k\neq i,j})
  ~.
\label{eq:thing}
\end{align}
We expand the covariances as a series in $\sqrt{t}$, assuming that they
are indeed expressible for short times using such an expansion:
\begin{align*}
\text{Cov}(z_{\sigma(1)},...,z_{\sigma(m)})
  & = \alpha_{\sigma(1),...,\sigma(m)}
  + \beta_{\sigma(1),...,\sigma(m)} \sqrt{t}
  + \gamma_{\sigma(1),...,\sigma(m)} t + O(t^{3/2})
  ~.
\end{align*}
As before, we substitute the above series expansion into Eqn.~\ref{eq:thing}
and match terms of $O(\frac{1}{t})$, $O(\frac{1}{\sqrt{t}})$, and $O(1)$ to
get:
\begin{align*}
0 & = -\frac{m}{2}\alpha_{\sigma(1),...,\sigma(m)}
  + \sum_{i,j} D_{i,j} \alpha_{\sigma(k):k\neq i,j} ~, \\
0 & = -\frac{m+1}{2} \beta_{\sigma(1),...,\sigma(m)}
  + \sum_{i,j} D_{i,j} \beta_{\sigma(k):k\neq i,j} ~,~\text{and} \\
0 & = -\gamma_{\sigma(1),...,\sigma(m)}
  -\frac{m}{2}\gamma_{\sigma(1),...,\sigma(m)}
  + \sum_{i,k} A_{ik} \alpha_{\sigma(k),\sigma(j):j\neq i}
  + \sum_{i,j} D_{ij} \beta_{\sigma(k):k\neq i,j}
	  ~.
\end{align*}
The base case is that, by definition, $\langle z\rangle = 0$ and $\langle
z^0\rangle =1$. This implies that $\beta_{\sigma(1),...,\sigma(m)}=0$ for all
lists $\{\sigma(i):i=1,...,m\}$. Since all moments are determined to at least
$O(t)$ by just the linearized version of the nonlinear Langevin equation and
since linear Langevin equations have Gaussian Green's functions, it follows that the
Green's function for the nonlinear Langevin equation is Gaussian to $O(t)$.
Some algebra shows that the variance of the linearized Langevin equation's
Green's function is:
\begin{align*}
\text{Var}(q(x|x')) = Dt + \frac{A(x')D+DA(x')^{\top}}{2}t^2 + O(t^3)
  ~.
\end{align*}
If $D$ is invertible, the conditional entropy is then:
\begin{align*}
H[X_{t+\tau}|X_t=x']
  & = \log \sqrt{2\pi e |\det (D \tau )|}
  + \frac{1}{2} \log \det
  \left(I+\frac{D^{-1} A(x') D + A(x')^{\top}}{2} \tau \right) + O(\tau^2) \\
  & = \log \sqrt{2\pi e |\det (D \tau)|}
  + \frac{1}{2}
  \text{tr} \left(\frac{D^{-1} A(x') D + A(x')^{\top}}{2} \tau \right)
  + O(\tau^2) \\
  & = \log \sqrt{2\pi e |\det (D\tau)|}
  + \frac{\text{tr}(A(x'))}{2} \tau  + O(\tau^2)
  ~.
\end{align*}
If the matrix $D$ is not invertible because $\det D = 0$, then we only have the leading order term in $t$ of the entropy $H[p]$ and we cannot draw any conclusions about the $O(1)$ term in any of the information anatomy quantities.  This becomes very clear by example in Appendix \ref{sec:linearLangevin}.

\section{Linear Langevin Dynamics with Noninvertible Diffusion Matrix}
\label{sec:linearLangevin}

If the stochastic differential equation is linear:
\begin{align}
\frac{dx}{dt} = A + Bx + \eta(t)
  ~,
\label{eq:linear}
\end{align}
where $\eta(t)$ is white noise $\langle \eta(t)\rangle = 0$ and
$\langle \eta(t) \eta(t')^{\top} \rangle = D \delta(t-t')$, then we can solve
Eqn.~\ref{eq:linear} in terms of $\eta(t)$ as:
\begin{align*}
\frac{dx}{dt} - B x & = A + \eta(t) \\
\frac{d}{dt}(e^{-Bt} x) & = e^{-Bt} A + e^{-Bt} \eta(t) \\
e^{-Bt} x(t) - x(0) & = \int_0^t e^{-B t'} A dt' + \int_0^t e^{-B t'} \eta(t') dt' 
  ~,
\end{align*}
yielding:
\begin{align*}
x(t) = e^{Bt} x(0) + \int_0^t e^{B(t-t')}A dt'
  + \int_0^t e^{B(t-t')} \eta(t') dt'
  ~.
\end{align*}
Since $\eta(t)$ is white, $x(t)$ is a Gaussian random variable with mean:
\begin{align*}
\langle x(t)\rangle = e^{Bt} x(0) + \int_0^t e^{B(t-t')}A dt'
\end{align*}
and variance:
\begin{align}
\text{Var}(x(t))
  & = \langle (x(t)-\langle x(t)\rangle) (x(t)-\langle x(t)\rangle)^{\top}\rangle
  \nonumber \\
  & = \left\langle \int_0^t e^{B(t-t')} \eta(t') dt'
  \int_0^{t} \eta(t'')^{\top} e^{B^{\top}(t-t'')} dt'' \right\rangle 
  \nonumber \\
  & = \int_0^t e^{B(t-t')} D e^{B^{\top}(t-t')} dt' 
  \nonumber \\
  & = \int_0^t e^{B t'} D e^{B^{\top} t'} dt'
  ~.
\label{eq:Var_linear}
\end{align}
Finally, we can also calculate the stationary probability distribution's
variance in several ways, but for now we simply define:
\begin{align*}
\Sigma = \lim_{t\rightarrow\infty} \text{Var}(x(t))
  ~.
\end{align*}
Since the Green's function is Gaussian for all time---not approximately in
the short time limit---and since the variance of this Gaussian does not depend
on the initial start point, we can calculate the conditional entropies
$H[X_{t}|X_0]$ via:
\begin{align}
H[X_{t}|X_0] = \frac{1}{2}\log (2\pi e |\det \text{Var}(x(t))|)
  ~.
\label{eq:condentropylinear}
\end{align}
The goal here is to calculate this quantity for small $t$ when the matrix $D$
is not invertible. We assume that it has the block matrix form:
\begin{align*}
D = \begin{pmatrix} 0 & 0 \\ 0 & D_{nn} \end{pmatrix}
  ~,
\end{align*}
where $D_{nn}^{\top} = D_{nn}$. Let $B$ have the corresponding block matrix
form:
\begin{align*}
B = \begin{pmatrix} B_{dd} & B_{dn} \\ B_{nd} & B_{nn} \end{pmatrix}
  ~.
\end{align*}
(Recall that the subscript $d$ stands for deterministic and the subscript $n$ stands for noisy.)  We can rewrite the variance in Eqn.~\ref{eq:Var_linear} as a power series in $t$:
\begin{align}
\text{Var}(x(t)) & = \int_0^t e^{B t'} D e^{B^{\top} t'} dt'
  \nonumber \\
  & = \int_0^t \left( \sum_{k=0}^{\infty} \frac{B^k}{k!} (t')^k \right) D 
  \left( \sum_{j=0}^{\infty} \frac{(B^{\top})^j}{j!} (t')^j \right) dt' 
    \nonumber \\
  & = \sum_{k,j=0}^{\infty} \frac{B^k D (B^{\top})^j}{k! j!} \int_0^t (t')^{k+j} dt' 
    \nonumber \\
  & = \sum_{k,j=0}^{\infty} \frac{B^k D (B^{\top})^j}{k! j!} \frac{t^{k+j+1}}{k+j+1} 
    \nonumber \\
  & = \sum_{m=1}^{\infty} \frac{t^m}{m!} \sum_{k=0}^{m-1} {m-1 \choose k} B^k D (B^{\top})^{m-1-k}
  ~.
\label{eq:var_powerseries}
\end{align}
Since we are concerned about the small-$t$ limit, we consider only with the
first few terms of this power series and, for reasons that will become clear,
we write all in block-matrix form. The first term, which is of $O(t)$,
is the usual:
\begin{align}
Q_1 & = D t \nonumber \\
  & = \begin{pmatrix} 0 & 0 \\ 0 & D_{nn} \end{pmatrix} t
  ~.
\label{eq:Q1}
\end{align}
The second term, of $O(t^2)$, has the form:
\begin{align}
Q_2 & = \frac{t^2}{2} (BD + DB^{\top}) \nonumber \\
  & = \frac{t^2}{2} \begin{pmatrix} 0 & B_{dn}D_{nn} \\ D_{nn} B_{dn}^{\top} & B_{nn} D_{nn} + D_{nn} B_{nn}^{\top} \end{pmatrix}
  ~.
\end{align}
The third term, of $O(t^3)$, has the form:
\begin{align}
Q_3 & = \frac{t^3}{6} (B^2 D+2BDB^{\top} + D(B^{\top})^2) \nonumber \\
  & = \frac{t^3}{6} \begin{pmatrix} 0 & B_{dd} B_{dn} D_{nn} \\ (B_{dd} B_{dn} D_{nn})^{\top}  & -- \end{pmatrix}
  + \frac{t^3}{3} \begin{pmatrix} B_{dn} D_{nn} B_{dn}^{\top} & B_{dn} D_{nn} B_{nn}^{\top} \\ B_{nn} D_{nn} B_{dn}^{\top} & --\end{pmatrix}
  ~.
\end{align}
We place a dash in the lower right block matrix entry since, as it turns out,
it does not matter for this calculation.
The fourth term, of $O(t^4)$, has the form:
\begin{align}
Q_4 & = \frac{t^4}{24} (B^3 D+3B^2DB^{\top} + 3BD(B^{\top})^2 + D(B^{\top})^3)
  \nonumber \\
  & = \frac{t^3}{6} \begin{pmatrix} (B_{dd}B_{dn}+B_{dn}B_{nn}) D_{nn} B_{dn}^{\top} & -- \\ -- & -- \end{pmatrix}
  + \frac{t^3}{6} \begin{pmatrix}  B_{dn} D_{nn} (B_{dd}B_{dn}+B_{dn}B_{nn})^{\top} & -- \\ -- & -- \end{pmatrix}
  ~.
\label{eq:Q4}
\end{align}
Similar to the $Q_3$ calculation, we care only about the upper left hand entry,
and so every other matrix entry can be ignored. Substituting
Eqns.~\ref{eq:Q1}-\ref{eq:Q4} into Eqn.~\ref{eq:var_powerseries}, we find that:
\begin{align}
\text{Var}(x(t)) = \begin{pmatrix} Q_{dd} & Q_{dn} \\ Q_{dn}^{\top} & Q_{nn}\end{pmatrix}
  ~.
\label{eq:variance2}
\end{align}
where:
\begin{align*}
Q_{nn} & = D_{nn} t + \frac{B_{nn} D_{nn} + D_{nn} B_{nn}^{\top}}{2} t^2 + O(t^3) \\
Q_{dn} & = \frac{B_{dn} D_{nn}}{2} t^2 + \frac{B_{dd} B_{dn} D_{nn} + 2B_{dn} D_{nn} B_{nn}^{\top}}{6} t^3  + O(t^4) \\
Q_{dd} & = \frac{B_{dn} D_{nn} B_{dn}^{\top}}{3} t^3 + \frac{(B_{dd}B_{dn} + B_{dn}B_{nn})D_{nn} B_{dn}^{\top}}{8} t^4 \\
  & \quad + \frac{B_{dn} D_{nn} (B_{dd}B_{dn} + B_{dn}B_{nn})^{\top}}{8} t^4 + O(t^5)
  ~.
\end{align*}
To find the determinant of the matrix in Eqn.~\ref{eq:variance2}, we use:
\begin{align}
\det \text{Var}(x(t))
  = \det Q_{nn} \det (Q_{dd} - Q_{dn} Q_{nn}^{-1} Q_{dn}^{\top} )
  ~.
\label{eq:detVar2}
\end{align}
Since $\det D_{nn} \neq 0$, $D_{nn}$ is invertible:
\begin{align}
\det Q_{nn} & = \det (D_{nn} t)
  \det (I+ \frac{D_{nn}^{-1} B_{nn} D_{nn}+ B_{nn}^{\top}}{2} t + O(t^2)) 
  \nonumber \\
  & = \det (D_{nn} t) (1+\text{tr}(B_{nn}) t + O(t^2)
  ~.
\label{eq:detQnn}
\end{align}
Again, we have used the fact that:
\begin{align*}
\text{tr}(D_{nn}^{-1} B_{nn} D_{nn} + B_{nn}^{\top})
  & = \text{tr}(B_{nn} D_{nn} D_{nn}^{-1})  + \text{tr}(B_{nn}^{\top}) \\
  & = 2\text{tr}(B_{nn})
  ~.
\end{align*}
And, since $D_{nn}$ is invertible, we can also write:
\begin{align*}
Q_{nn}^{-1}
  & = D_{nn}^{-1} t^{-1}
  \left( 1+\frac{D_{nn}^{-1}B_{nn}D_{nn} + B_{nn}^{\top}}{2} t \right)^{-1}
  + O(t) \\
  & = D_{nn}^{-1} t^{-1}
  \left( 1-\frac{D_{nn}^{-1}B_{nn}D_{nn} + B_{nn}^{\top}}{2} t \right)
  + O(t)
  ~.
\end{align*}
Then:
\begin{align*}
Q_{dn} Q_{nn}^{-1} Q_{dn}^{\top}
  & = \left( \frac{B_{dn} D_{nn}}{2} t^2 + \frac{B_{dd} B_{dn} D_{nn}
  + 2B_{dn} D_{nn} B_{nn}^{\top}}{6} t^3 \right) \\
  & \quad\quad \times \frac{D_{nn}^{-1}}{t}
  \left(1-\frac{D_{nn}^{-1}B_{nn}D_{nn} + B_{nn}^{\top}}{2} t \right) \\
  & \quad\quad \times \left(\frac{B_{dn} D_{nn}}{2} t^2
  + \frac{B_{dd} B_{dn} D_{nn} + 2B_{dn} D_{nn} B_{nn}^{\top}}{6} t^3
  \right)^{\top}
  + O(t^5)
  ~.
\end{align*}
With some algebra, this becomes:
\begin{align*}
Q_{dn} Q_{nn}^{-1} Q_{dn}^{\top}
  = \frac{B_{dn}D_{nn}B_{dn}^{\top}}{4}t^3
  & + \frac{B_{dd}B_{dn} D_{nn} B_{dn}^{\top} + 2B_{dn} D_{nn} B_{nn}^{\top} B_{dn}^{\top}}{12} t^4 \\
  & + \frac{B_{dn} D_{nn} B_{dn}^{\top} B_{dd}^{\top} + 2B_{dn} B_{nn} D_{nn} B_{dn}^{\top}}{12} t^4 \\
  & -\frac{B_{dn}D_{nn}^{-1} B_{nn} D_{nn}^2 B_{dn}^{\top} + B_{dn} B_{nn}^{\top} D_{nn} B_{dn}^{\top}}{8} t^4
  + O(t^5)
  ~.
\end{align*}
Assume that $B_{dn}D_{nn} B_{dn}^{\top}$ is invertible; i.e.,
$\det (B_{dn} D_{nn} B_{dn}^{\top})\neq 0$. Therefore:
\begin{align}
F & = \det (Q_{dd} - Q_{dn} Q_{nn}^{-1} Q_{dn}^{\top} ) \nonumber \\
  & = \det \Big( \frac{B_{dn}D_{nn}B_{dn}^{\top}}{12} t^3
  + \frac{(B_{dd}B_{dn}-B_{dn}B_{nn})D_{nn}B_{dn}^{\top}}{24} t^4 \nonumber \\
  & + \frac{B_{dn} D_{nn} (B_{dd}B_{dn}-B_{dn}B_{nn})^{\top}}{24} t^4
  \nonumber \\
  & -\frac{B_{dn}D_{nn}^{-1} B_{nn} D_{nn}^2 B_{dn}^{\top}
  + B_{dn} B_{nn}^{\top} D_{nn} B_{dn}^{\top}}{8} t^4 \Big) + O(t^2)
  \nonumber \\
  & = \det \left(\frac{B_{dn}D_{nn}B_{dn}^{\top}}{12} t^3)
  (1+\text{tr}(M) t \right)
  + O(t^2)
  ~,
\label{eq:determinantF}
\end{align}
where
\begin{align*}
M & = \frac{(B_{dn}D_{nn}B_{dn}^{\top})^{-1}B_{dn}D_{nn}
  (B_{dd}B_{dn}-B_{dn}B_{nn})^{\top}}{2}
  + \frac{(B_{dn}D_{nn}B_{dn}^{\top})^{-1}
  (B_{dd}B_{dn}-B_{dn}B_{nn})D_{nn}B_{dn}^{\top}}{2} \\
  & \quad\quad - \frac{3(B_{dn}D_{nn}B_{dn}^{\top})^{-1} B_{dn}D_{nn}^{-1}
  B_{nn} D_{nn}^2 B_{dn}^{\top}}{2}
  - \frac{3(B_{dn}D_{nn}B_{dn}^{\top})^{-1} B_{dn} B_{nn}^{\top}
  D_{nn} B_{dn}^{\top}}{2}
  ~.
\end{align*}
Substituting Eqns.~\ref{eq:detQnn} and \ref{eq:determinantF} into
Eqn.~\ref{eq:detVar2} and substituting that into
Eqn.~\ref{eq:condentropylinear}, we have the conditional entropy:
\begin{align*}
H[X_t|X_0]
  = & \log \sqrt{|\det D_{nn}|} + \log\sqrt{|\det
  B_{dn}D_{nn}B_{dn}^{\top}|} + \frac{3m+n}{2} \log t \\
  & - m \log\sqrt{12} + \frac{\text{tr}(M) t +\text{tr}(B_{nn}) t}{2}
  + \log\sqrt{2\pi e} + O(t^2)
  ~,
\end{align*}
where $m=\text{dim}(B_{dd})$ and $n=\text{dim}(B_{nn})$. Suppose that $B_{dn}
B_{dn}^{\top}$ is invertible. Then, some algebra not shown here reveals:
\begin{align*}
\text{tr}(M)
  & = \text{tr}(B_{dd})-\text{tr}(B_{nn}) - 3\text{tr}(B_{nn}) \\
  & = \text{tr}(B_{dd})-4\text{tr}(B_{nn})
  ~.
\end{align*}
For this special case, the conditional entropy is:
\begin{align*}
H[X_t|X_0]
  = & \log|\det D_{nn}| + \log|\det B_{dn} D_{nn} B_{dn}^{\top}|
  + \log\sqrt{2\pi e} \\
  & + \frac{3m+n}{2} \log t - m \log\sqrt{12}
  + \frac{1}{2} \left( \text{tr}(B_{dd})-3\text{tr}(B_{nn}) \right) t + O(t^2)
  ~.
\end{align*}

\section{Time-local Predictive Information}
\label{sec:TiPi}

Information anatomy measures will have broad application to monitoring and
guiding the behavior of adaptive autonomous agents. Practically, information
anatomy gives a suite of semantically distinct kinds of information
\cite{Crut10a,Jame11a} that is substantially richer and structurally more
incisive than simple uses of Shannon mutual information that implicitly assume
there is only a single kind of (correlational) information.  For example, it is
reasonable to hypothesize that biological sensory systems are optimized to
transmit with high fidelity information that is predictively useful about
stimuli or environmental organization. In such a setting, the bound information
quantifies how much predictability is lost if one has extracted the full
predictable information $\EE$ from the past, but chooses to ignore the present
$H(\MS_0)$.  Along these lines, the \emph{time-local predictive
information}\footnote{For clarity, we must address a persistently misleading
terminology at use here, since it is critical to correctly interpreting
the benefits of information-theoretic analyses. The proposed measure is a special case of bound
information $\bmu$. Recall that both $\bmu$ and the excess entropy $\EE$
capture the amount of information in the future that is \emph{predictable}
\cite{Crut01a,Jame11a} and \emph{not} that which is \emph{predictive}. The
latter is the amount of information that must be stored to optimally predict
and this is given by the statistical complexity $\Cmu$.  And so, when we use the
abbreviation TiPi, we mean the \emph{time-local predictable
information}---information the agent immediately sees as advantageous. } (TiPi)
was recently proposed as a quantity that agents maximize in order to access
different behavioral modes when adapting to their environment \cite{Mart13a}.

In fact, \cite{Mart13a} does a calculation very similar to the ones above,
considering discrete-time stochastic dynamics of the form:
\begin{align*}
x_t = \phi(x_{t-1}) + \eta_t
\end{align*}
and calculating the TiPi:
\begin{align}
I^{T}[X_{t};X_{t-1}] \equiv I[X_t;X_{t-1}|X_{t-T} = x_{t-T}] ~,
\label{eq:TiPiDefn}
\end{align}
with fixed $T > 1$. The motivation being that, whatever the history prior to
$t-T$, the agent knows the environment state $x_{t-T}$ then. However, from that
time forward the agent, making no further observations, is ignorant. The
stochastic dynamics then models the evolution of that ignorance from the given
state to a distribution of states at $t-1$ and then at $t$, taking into
account only the model $\phi$ the agent has learned or is given. They report
that TiPi is the difference between state information and noise entropy:
\begin{align}
I^{T}[X_{t};X_{t-1}] = \tfrac{1}{2}\ln |\det \Sigma|
	- \tfrac{1}{2}\ln |\det D|
	~,
\label{eq:tipi}
\end{align}
where:
\begin{align}
D & = \langle \eta \eta^{\top}\rangle ~, \nonumber \\
  \Sigma & = \sum_{k=1}^{T} L(x_{t-k}) D L(x_{t-k})^{\top} ~,
  \label{eq:tipi2} \\
  \left(L(x)\right)_{ij} & = \frac{\partial\phi_i(x)}{\partial x_j}
  \nonumber
  ~,
\end{align}
and
\begin{equation}
L^{(k)}(t-1) = \prod_{m=1}^{k} L(x_{t-m})
  \nonumber
  ~,
\end{equation}
with $L^{(0)} = I$.

Since $\Sigma$ depends on the states between times $t-T$ and $t-1$, the TiPi
expression in Eqn.~\ref{eq:tipi} also depends on the states between times $t-T$
and $t-1$. The TiPi definition in Eqn.~\ref{eq:TiPiDefn} does not.
%There is a concern, then, with the
%calculation as presented in \cite{Mart13a}. The conditional mutual information
%$I[X;Y|Z]$ is $\int p(x,y,z) \log \frac{p(x,y|z)}{p(x|z) p(y|z)} dx dy dz$
%and, obviously, does not depend on any specific realizations of $X$, $Y$, or
%$Z$. Likewise, the conditional mutual information $I[X_t;X_{t-1}|X_{t-T}]$
%cannot depend on specific values of $x_{t-1}$ and $x_{t-T}$. Also, the
%conditional mutual information $I[X_t;X_{t-1}|X_{t-T}]$ cannot depend on
%specific values of $x_{t'}$ for $t' = t-2,...,t-T+1$, as the conditional
%probability distribution $p(x_t,x_{t-1}|x_{t-T})$ already involves
%marginalizing over such $x_{t'}$.
Thus, even though the numerical results of \cite{Mart13a} are quite
interesting, the quantity that the behavioral agents there were maximizing was
\textit{not} the stated conditional mutual information.

To address this concern and explore informational adaptation hypotheses, let's consider
alternatives. If desired, for example, one could define an averaged TiPi as:
\begin{align*}
I^{T}_1[X_t;X_{t-1}] & \equiv I[X_t;X_{t-1}|X_{t-T}] \\
  & = H[X_t|X_{t-T}] - H[X_t|X_{t-1},X_{t-T}]
  ~.
\end{align*}
Or, one could define TiPi to be:
\begin{align*}
I^{T}_2[X_t;X_{t-1}] & \equiv H[X_t|X_{t-T}=x_{t-T}]- H[X_t|X_{t-1}=x_{t-1}]
  ~,
\end{align*}
so that it depends on both $x_{t-T}$ and $x_{t-1}$.

Even with these modifications, Eqn.~\ref{eq:tipi} still cannot be a general
expression for TiPi since it depends on measurements at intermediate times that must be marginalized out of the conditional probability distribution with which we are calculating the mutual information.

Moving to discrete time with a small discretization time, let's find
expressions for all three:
\begin{align*}
I^{N}[X_t;X_{t-\tau}]
  & = I[X_t;X_{t-\tau}|X_{t-N\tau}=x_{t-N\tau}]
%  \label{eq:TiPi2}
  \\
I_1^{N}[X_t;X_{t-\tau}]
  & = I[X_t;X_{t-\tau}|X_{t-N\tau}]
%  \label{eq:TiPi1}
  \\
I_2^{N}[X_t;X_{t-\tau}]
  & = H[X_t|X_{t-N\tau}=x_{t-N\tau}]- H[X_t|X_{t-\tau}=x_{t-\tau}]
%  \label{eq:TiPi3}
  ~.
\end{align*}
Suppose that the underlying dynamical system is a nonlinear Langevin equation with invertible diffusion matrix and an analytic potential function $U_{\theta}$ parametrized by $\theta$:
\begin{align*}
\frac{dx}{dt} & = -D\nabla U_{\theta}(x) + \eta(t) ~,
\end{align*}
with white noise---$\langle\eta(t)\rangle = 0$ and
$\langle \eta(t)\eta(t')^{\top} \rangle = D \delta(t-t')$.
Following the argument used in Section \ref{sec:first-order}:
\begin{align*}
H[X_t|X_{t-N\tau}] & = \log\sqrt{2\pi e (N\tau)^n |\det D|} - \frac{N\tau}{2} \int \nabla\cdot (D\nabla U_{\theta}(x)) P(X_{t-N\tau}=x) dx
  + O((N\tau)^2) ~, \\
H[X_t|X_{t-N\tau}=x_{t-N\tau}] & = \log\sqrt{2\pi e (N\tau)^n |\det D|} - \frac{N\tau}{2}\nabla\cdot(D\nabla U_{\theta}(x))|_{x=x_{t-N\tau}}
  + O((N\tau)^2) ~, \\
H[X_t|X_{t-\tau}] & = \log\sqrt{2\pi e \tau^n |\det D|} - \frac{\tau}{2} \int \nabla\cdot (D\nabla U_{\theta}(x)) P(X_{t-\tau}=x) dx
  + O(\tau^2) ~, ~\text{and} \\
H[X_t|X_{t-\tau}=x_{t-\tau}] & = \log\sqrt{2\pi e \tau^n |\det D|} - \frac{\tau}{2} \nabla\cdot (D\nabla U_{\theta}(x))|_{x_{t-\tau}}
  + O(\tau^2)
  ~.
\end{align*}
These formulae lead to the following expressions for the TiPi alternatives:
\begin{align}
I^{N}[X_t;X_{t-\tau}] & = n\log\sqrt{N} - \frac{N\tau}{2} \nabla\cdot (D\nabla
U_{\theta}(x))_{x=x_{t-N\tau}}\nonumber \\
  & + \frac{\tau}{2} \int \nabla\cdot (D\nabla U_{\theta}(x))
  P(X_{t-\tau}=x|X_{t-N\tau}=x_{t-N\tau}) dx + O((N\tau)^2)
  \label{eq:TiPi1_calc} ~, \\
I_1^{N}[X_t;X_{t-\tau}] & = n\log\sqrt{N} - \frac{N\tau}{2}\int \nabla\cdot (D\nabla U_{\theta}(x)) P(X_{t-N\tau}=x) dx \nonumber \\
  & + \frac{\tau}{2} \int \nabla\cdot (D\nabla U_{\theta}(x)) P(X_{t-\tau}=x)
  dx + O((N\tau)^2) \label{eq:TiPi0_calc} ~, ~\text{and} \nonumber \\
I_2^{N}[X_t;X_{t-\tau}] & = n\log\sqrt{N} - \frac{N\tau}{2} \nabla \cdot (D\nabla U_{\theta}(x))|_{x_{t-N\tau}} \nonumber \\
  & + \frac{\tau}{2} \nabla \cdot (D\nabla U_{\theta}(x))|_{x_{t-\tau}} + O((N\tau)^2)
  \nonumber ~.
\end{align}
Maximizing these with respect to $\theta$ has a different effect on the action
policy. Maximizing the original TiPi $I^N[X_t;X_{t-\tau}]$ leads the agent to
alter the landscape so that it is driven into unstable regions. Maximizing
the averaged TiPi $I_1^N[X_t;X_{t-\tau}]$ leads to a flattening of the
potential landscape. And, the effect of maximizing $I_2^N[X_t;X_{t-\tau}]$ is not yet clear.

% In the following commented out text, JPC did not yet swap I^N and I_1^N !!!
%For intuition purposes, suppose that the drift is itself the gradient of a
%potential function, $\mu = -\nabla U_{\theta}(x)$, where $\theta$ are
%parameters which the agent can alter to change the shape of the potential
%landscape.  An agent trying to maximize $I^{N}[X_t;X_{t-\tau}]$ will flatten
%the potential function.  An agent trying to maximize $I_1^{N}[X_t;X_{t-\tau}]$
%will try to move to regions of local maxima (unstable equilibria) while
%flattening the potential function, which are two competing desires; the
%balance between these two desires is governed by $N$.  An agent trying to
%maximize $I_1^{N}[X_t;X_{t-\tau}]$ will try to consistently move to local
%maxima while turning previously visited locations into local minima, which
%could lead to interesting dynamic behavior. Further, pursuing learning rules
%for changing $\theta$ so as to maximize these quantities would also be interesting.

Not surprisingly, when $N$ is small, we recover the result that maximizing
$I^N[X_t;X_{t-\tau}]$ has the same effect on the potential landscape as
maximizing the TiPi in \cite{Mart13a} when $T=2$. Though the model there is
set up for a discrete-time analysis, it is natural to suppose that adaptive
agents in an environment move according to continuous-time equations, but
receive sensory signals in a discrete-time manner. Equating notation used
here and there:
\begin{align*}
\phi(x) = x-D\nabla U(x)\tau 
\end{align*}
gives:
\begin{align*}
L^{(1)} = I - A(x_{t-\tau}) \tau ~,
\end{align*}
where $A_{ij} = \partial (D\nabla U(x))_i / \partial x_j$. When $N=2$,
substituting this into Eqn.~\ref{eq:tipi2} yields:
\begin{align*}
\Sigma & = D\tau + L^{(1)} D\tau (L^{(1)})^{\top} \\
   & = 2D\tau  - (D A(x_{t-\tau})^{\top} + A(x_{t-\tau}) D)\tau^2 + O(\tau^3)
  ~.
\end{align*}
This then gives, upon substitution into Eqn.~\ref{eq:tipi}:
\begin{align*}
\frac{1}{2}\log|\Sigma|-\frac{1}{2}\log |D\tau|
  & = n \log \sqrt{2} - \text{tr}(A(x_{t-\tau})) \tau  + O(\tau^2) \\
  & = n\log\sqrt{2} - \nabla\cdot (D\nabla U(x))|_{x_{t-\tau}} \tau + O(\tau^2)
  ~.
\end{align*}
The above expression is identical to that in Eqn.~\ref{eq:TiPi1_calc} for all
practical purposes, as derivatives of the two with respect to $\theta$ are
identical up to an unimportant multiplicative constant to subleading order in
$\tau$. Therefore, for $T=2$ many of the qualitative conclusions from numerical
simulations are likely to carry over when Eqn.~\ref{eq:TiPi1_calc} is used as
the objective function.

Finally, the difference in how these quantities were calculated is interesting
to us. For instance, was the series expansion for the coefficients of the
moments of the Green's function in Appendix \ref{sec:Greens} actually
necessary?  Could we have used an Ito discretization scheme to write
$x_{t+\Delta t}$ in terms of $x_{t-\Delta t}$ and noise terms, and use that
expression to evaluate $b_{\mu}$?  This is related to the approach taken in
\cite{Mart13a}.
%The most worrisome part about this appealing approach is
%that by discretizing a nonlinear Langevin equation on the same time scale as
%the measurement time scale, rather than discretizing on a time scale much
%smaller than the measurement time resolution, one might erroneously throw away
%terms.
However, the answer obtained using the moment series expansions is a
factor of two different than would have been obtained such a discretization
scheme. And, by keeping track of the order of the approximation errors
in Appendix \ref{sec:Greens}, we found that these formulae for both bound
information and TiPi would only hold for invertible diffusion matrices.  As
suggested by Appendix \ref{sec:linearLangevin}, our estimates for such
conditional mutual informations change qualitatively when the diffusion matrix
is not invertible. And that, in turn, may be relevant to environments that are
hidden Markovian---settings for which the agent's sensorium does not directly
report the environmental states.

\end{document}